\documentclass[aps,prl,letterpaper,showkeys,showpacs,twocolumn,nofootinbib,preprintnumbers,longbibliography]{revtex4-1}

\usepackage{amsmath,amssymb,lmodern,amsfonts} 
\usepackage{graphicx}
\usepackage{hyperref}
\usepackage{color}
\usepackage{natbib}
\usepackage[caption=false]{subfig}

\begin{document}

\preprint{PI/UAN-2017-615FT}

\title{Non-Abelian $S$-term dark energy and inflation}

\author{Yeinzon Rodr\'iguez}
\email{yeinzon.rodriguez@uan.edu.co}
\affiliation{Centro de Investigaciones en Ciencias B\'asicas y Aplicadas, Universidad Antonio Nari\~no, \\ Cra 3 Este \# 47A-15, Bogot\'a D.C. 110231, Colombia}
\affiliation{Escuela  de  F\'isica,  Universidad  Industrial  de  Santander, \\ Ciudad  Universitaria,  Bucaramanga  680002,  Colombia}
\affiliation{Simons Associate at The Abdus Salam International Centre for Theoretical Physics, \\ Strada Costiera 11, I-34151, Trieste, Italy}

\author{Andr\'es A. Navarro}
\email{andres.navarro@ustabuca.edu.co}
\affiliation{Escuela  de  F\'isica,  Universidad  Industrial  de  Santander, \\ Ciudad  Universitaria,  Bucaramanga  680002,  Colombia}
\affiliation{Departamento de Ciencias B\'asicas,  Universidad Santo Tom\'as, \\ Cra 17 \# 9-27,  Bucaramanga  680002,  Colombia}


\begin{abstract}
We study the role that a cosmic triad in the generalized $SU(2)$ Proca theory, specifically in one of the pieces of the Lagrangian that involves the symmetric version $S_{\mu \nu}$ of the gauge field strength tensor $F_{\mu \nu}$, has on dark energy and primordial inflation.  Regarding dark energy, the triad behaves asymptotically as a couple of radiation perfect fluids whose energy densities are negative for the $S$ term but positive for the Yang-Mills term.  This leads to an interesting dynamical fine-tuning mechanism that gives rise to a combined equation of state parameter $\omega \simeq -1$ and, therefore, to an eternal period of accelerated isotropic expansion for an ample spectrum of initial conditions.  Regarding primordial inflation, one of the critical points of the associated dynamical system can describe a prolonged period of isotropic slow-roll inflation sustained by the $S$ term.  This period ends up when the Yang-Mills term dominates the energy density leading to the radiation dominated epoch.  Unfortunately, in contrast to the dark energy case, the primordial inflation scenario is strongly sensitive to the coupling constants and initial conditions.  The whole model, including the other pieces of the Lagrangian that involve $S_{\mu \nu}$, might evade the recent strong constraints coming from the gravitational wave signal GW170817 and its electromagnetic counterpart GRB 170817A.
\end{abstract}



\maketitle

{\it Introduction} - The vector sector of gauge field theories is built from the gauge field strength tensor $F_{\mu \nu}$, its Hodge dual $\tilde{F}_{\mu \nu}$, and, if the gauge symmetry is spontaneously broken, from the vector field $A_\mu$ \cite{Weinberg:1996kr}.  Generalized Proca theories have taught us that, when the gauge symmetry is explicitly broken, the vector sector of these theories is also built from the symmetric version $S_{\mu \nu}$ of $F_{\mu \nu}$ \cite{Jimenez:2016isa,Allys:2016jaq} (see also Refs. \cite{Rodriguez:2017ckc,Heisenberg:2017mzp}).  The cosmological implications of $F_{\mu \nu}$, $\tilde{F}_{\mu \nu}$,  and $A_\mu$ have been well investigated in the literature (see, for instance Refs. \cite{Maleknejad:2012fw,Soda:2012zm,Dimopoulos:2011ws}) but little has been said about $S_{\mu \nu}$.  In this paper, we study the cosmological implications of a cosmic triad \cite{ArmendarizPicon:2004pm} in the vector-tensor Horndeski theory, also called the theory of vector Galileons, endowed with a global $SU(2)$ symmetry.  In particular, we analyse the Yang-Mills Lagrangian together with $\mathcal{L}^1_4 \subset \mathcal{L}_4$, it being one of the pieces of the generalized $SU(2)$ Proca Lagrangian \cite{Allys:2016kbq} that contains contractions of two $S_{\mu \nu}$.  We have found an asymptotic behaviour in which the cosmic triad under $\mathcal{L}^1_4$ behaves as an almost  radiation-like perfect fluid with negative energy density and pressure whose absolute values matches almost precisely those of the radiation perfect fluid coming from the same cosmic triad under the Yang-Mills Lagrangian. The system exhibits an interesting dynamical fine-tuning mechanism which results in a combined equation of state parameter $\omega \simeq -1$ and, therefore, in an eternal isotropic inflationary period;  this makes of this model an ideal candidate to explain the dark energy.  We have also explored the dynamical system associated to this model and we have found that one of the critical points may correspond to a prolonged period of isotropic slow-roll accelerated expansion.  This is a saddle point, i.e., it represents a transient state of the dynamical system so that the inflationary period comes naturally to an end, this being replaced by a radiation dominated period by virtue of the Yang-Mills Lagrangian;  this model would be an ideal candidate to explain the primordial inflation were it not for the necessary judicious choosing of initial conditions and parameters in the action.  The purpose of this paper is to isolate and understand the cosmological implications of $\mathcal{L}^1_4$ despite of being apparently strongly constrained \cite{Baker:2017hug,Creminelli:2017sry,Sakstein:2017xjx,Ezquiaga:2017ekz,Wang:2017rpx} by the recent observation of the gravitational wave signal GW170817 \cite{TheLIGOScientific:2017qsa} and its electromagnetic counterpart GRB 170817A \cite{GBM:2017lvd,Monitor:2017mdv} \footnote{We say ``apparently strongly constrained'' because there does not exist a formal proof of it.  The analyses so far done are for a scalar Galileon \cite{Creminelli:2017sry,Sakstein:2017xjx,Ezquiaga:2017ekz,Wang:2017rpx} and for the generalized Proca action for an Abelian vector field \cite{Baker:2017hug}.}.  The purpose is reasonable since the generalized $SU(2)$ Proca Lagrangian contains $\mathcal{L}_4 \equiv \alpha \mathcal{L}_4^1 + \kappa \mathcal{L}_4^2 + \lambda \mathcal{L}_4^3$, $\alpha, \kappa, \lambda$ being constants, where a relation between $\alpha$ and $\kappa$ might be established so that the gravitational waves speed matches that of light\footnote{The cosmological implications of $\mathcal{L}_4^3$ were reported in Ref. \cite{Rodriguez:2017ckc}.  For its own existence, this parity-violating term requires not only at least one non-vanishing temporal component but also a non-orthogonal configuration for the triad, potentially generating anisotropies in the expansion in conflict with observations.}.  In such a scenario, although $\alpha,\kappa \neq 0$ in principle, $\kappa$ being a function of $\alpha$, it might happen the cosmological implications of $\mathcal{L}^1_4$ are not counterbalanced by those of $\mathcal{L}^2_4$.  We will analize such a scenario and the whole cosmological implications of $\alpha \mathcal{L}_4^1 + \kappa \mathcal{L}_4^2 \subset \mathcal{L}_4$ in a forthcoming publication.

{\it Generalized Proca theories and the cosmic triad} - Generalized Proca theories are built following the same construction idea of the Galileon-Horndeski theories \cite{Rodriguez:2017ckc,Deffayet:2013lga}.  Whatever choices Nature had to define the action, once the field content and the symmetries were decided, all of them must comply with a Hamiltonian bounded from below.  And this may be possible, according to Ostrogradski \cite{ostro}, if the dynamical field equations are, at most, second order in space-time derivatives.  If the latter condition were not satisfied, the system would generically enter in a severe instability, called Ostrogradski's, both at the classical and quantum levels \cite{Woodard:2006nt,Woodard:2015zca}.  The traditional approach to construct such theories is by employing scalar fields as the field content \cite{Horndeski:1974wa,Nicolis:2008in,Deffayet:2009wt,Deffayet:2009mn,Deffayet:2011gz,Kobayashi:2011nu}.  Nothing significantly new, compared to the usual canonical kinetic term, is obtained when employing, instead, an Abelian gauge field \cite{Horndeski:1976gi,Deffayet:2013tca}.  Hence, having new phenomenology requires no longer invoking gauge symmetries, i.e., it requires a generalization to the Proca action.  Such a generalization was performed in Refs. \cite{Heisenberg:2014rta,Tasinato:2014eka,Hull:2015uwa,Allys:2015sht,Jimenez:2016isa,Allys:2016jaq} where it was recognized that, besides $F_{\mu \nu}$ and its Hodge dual $\tilde{F}_{\mu \nu}$, the action is also defined in terms of $A_\mu$ and the symmetric version $S_{\mu \nu}$ of $F_{\mu \nu}$:  $S_{\mu \nu} \equiv \nabla_\mu A_\nu + \nabla_\nu A_\mu$.  The application of all these ideas to non-Abelian theories culminated in the construction of the generalized $SU(2)$ Proca theory \cite{Allys:2016kbq} (see also Ref. \cite{Jimenez:2016upj}).  An interesting aspect of this theory  is the explicit violation of the $SU(2)$ gauge symmetry which allows a mass term and its generalizations written in terms of the non-Abelian versions of $A_\mu$, $F_{\mu \nu}$, $\tilde{F}_{\mu \nu}$, and $S_{\mu \nu}$.  Another interesting aspect is the global character of the $SU(2)$ symmetry which might play an important role in particle physics\footnote{Global continuous symmetries are important in particle physics, say, for example, in the solution to the strong CP problem via the spontaneous breaking of the $U(1)$ global symmetry imposed by the Peccei-Quinn mechanism \cite{Weinberg:1996kr}.}.  A third interesting aspect is the possibility of using a cosmic triad \cite{ArmendarizPicon:2004pm}, a set of three vector fields mutually orthogonal and of the same norm, which corresponds to an invariant configuration both under $SU(2)$, for the field space, and $SO(3)$, for the physical space, in agreement with the local homomorphism between these two groups.  The cosmic triad configuration has been employed before \cite{Maleknejad:2011jw,Maleknejad:2011sq,Adshead:2012kp,Nieto:2016gnp,Adshead:2016omu,Adshead:2017hnc,Davydov:2015epx} and, at least in the Gauge-flation scenario \cite{Maleknejad:2011jw,Maleknejad:2011sq}, its naturalness has been shown in the sense that it is an attractor in a more general anisotropic setup \cite{Maleknejad:2011jr}.  The cosmological implications of the generalized Proca theory for an Abelian vector field have been recently studied \cite{deFelice:2017paw,DeFelice:2016uil,DeFelice:2016yws,Tasinato:2014mia,Tasinato:2014eka} but always working with a time-like vector field so that the spatial components are chosen to vanish, avoiding this way disastrous anisotropies\footnote{An exception is the model studied in Ref. \cite{Emami:2016ldl} where a triad of space-like Abelian vector fields is considered so that the temporal components are chosen to vanish.  The results of this work are very interesting despite the unnaturalness of the triad configuration when there is no an underlying global $SU(2)$ symmetry.}.  In contrast, the isotropic configuration  provided by the cosmic triad, although the latter is composed of vector fields that inherently define privileged directions, is amply favoured by cosmological observations.   It is the purpose of this paper to focus on the spatial components of a triad of space-like vector fields.

{\it The non-Abelian $S$ terms and the considered model} - The Lagrangian of the generalized $SU(2)$ Proca theory is composed of several pieces that are described in Eqs. (96) - (99) of Ref. \cite{Allys:2016kbq}.  Of particular importance is $\mathcal{L}_4$ which is characterized by the two first-order covariant space-time derivatives of $A_\mu$ that each of its terms contain (except for the non-minimal coupling to gravity terms):
\begin{equation}
\mathcal{L}_4 \equiv \alpha \mathcal{L}_4^1 + \kappa \mathcal{L}_4^2 + \lambda \mathcal{L}_4^3 \,,
\end{equation}
with $\alpha,\kappa,\lambda \in \mathbb{R}$ and where\footnote{The difference between our $\mathcal{L}_4^1$ and that in Ref. \cite{Allys:2016kbq} is ${[(A_b \cdot A^b) G^{\mu a}_{\;\; \nu} G^{\;\; \nu}_{\mu a} + 2(A_a \cdot A_b) G^{\mu a}_{\;\; \nu} G^{b \; \nu}_\mu]/4}$.  Likewise, the difference between our $\mathcal{L}_4^2$ and that in Ref. \cite{Allys:2016kbq} is ${[(A_a \cdot A_b) G^{\mu a}_{\;\; \nu} G^{b \; \nu}_\mu - 2 (A^{\mu a} A^{\nu b}) (G^{\;\; \rho}_{\mu a} G_{\nu \rho b} - G^{\;\; \rho}_{\nu a} G_{\mu \rho b})]/4}$.  These differences formally belong to $\mathcal{L}_2 \equiv \mathcal{L}_2 (A^a_\mu, G^a_{\mu \nu}, \tilde{G}^a_{\mu \nu})$ in Eq. (96) of Ref. \cite{Allys:2016kbq}.}
\begin{eqnarray}
\mathcal{L}_4^1 &\equiv& \frac{1}{4}(A_b \cdot A^b) \left[ S^{\mu a}_\mu S^\nu_{\nu a} - S^{\mu a}_\nu S^\nu_{\mu a} + A_a \cdot A^a R \right] \nonumber \\
&& + \frac{1}{2} (A_a \cdot A_b) \left[S^{\mu a}_\mu S^{\nu b}_\nu - S^{\mu a}_\nu S^{\nu b}_\mu + 2 A^a \cdot A^b R\right] \,, \\
\mathcal{L}_4^2 &\equiv& \frac{1}{4}(A_a \cdot A_b) \left[S^{\mu a}_\mu S^{\nu b}_\nu - S^{\mu a}_\nu S^{\nu b}_\mu + A^a \cdot A^b R\right]  \nonumber \\
&& + \frac{1}{2} (A^{\mu a} A^{\nu b}) \Big[S^\rho_{\mu a} S_{\nu \rho b} - S^\rho_{\nu a} S_{\mu \rho b} - A^\rho_a A^\sigma_b R_{\mu\nu\rho\sigma} \nonumber \\
&& - \left(\nabla^\rho A_{\mu a} \right)\left(\nabla_\rho A_{\nu b} \right) + \left(\nabla^\rho A_{\nu a} \right)\left(\nabla_\rho A_{\mu b} \right)\Big] \,, \label{riemann} \\
\mathcal{L}_4^3 &\equiv& \tilde{G}_{\mu\sigma}^b A^\mu_a A_{\nu b}  S^{\nu\sigma a} \,.
\end{eqnarray}
In the previous expressions, gauge indices run from 1 to 3 and are represented by Latin letters, space-time indices run from 0 to 3 and are represented by Greek letters, $R$ is the Ricci scalar, $R_{\mu \nu \rho \sigma}$ is the Riemann tensor, $G_{\mu \nu}^a$ is the Abelian version of $F_{\mu \nu}^a$:
\begin{equation}
G_{\mu \nu}^a \equiv \nabla_\mu A_\nu^a - \nabla_\nu A_\mu^a \,,
\end{equation}
$\tilde{G}_{\mu \nu}^a$ is the Hodge dual of $G_{\mu \nu}^a$, and $S_{\mu \nu}^a$ is the symmetric version of $G_{\mu \nu}^a$:
\begin{equation}
S^a_{\mu \nu} \equiv \nabla_\mu A^a_\nu + \nabla_\nu A^a_\mu \,.
\end{equation}
It is very important to notice that the third line of $\mathcal{L}_4^2$, formed by products of two first-order covariant space-time derivatives of $A_\mu$, cannot be written either in terms of $F_{\mu \nu}^a$, $\tilde{F}_{\mu \nu}^a$, or $S_{\mu \nu}^a$, this line being a specific term to the non-Abelian nature of the theory \cite{Allys:2016kbq}.  As such, it vanishes in the Abelian case so that $\mathcal{L}_4^1 + \mathcal{L}_4^2$ reduces to $- A^2 [(S_\mu^\mu)^2 - S_\rho^\sigma S_\sigma^\rho] + \frac{1}{4} A^4 R$ which is part of the corresponding $\mathcal{L}_4$ in the generalized Proca theory for an Abelian vector field \cite{Rodriguez:2017ckc}.  This is the reason why we will denote $\mathcal{L}_4^1$ and $\mathcal{L}_4^2$ as the non-Abelian $S$ terms.  In this paper, we will analyse the cosmological consequences of the non-Abelian $S$ term in the action
\begin{equation}
\mathcal{S} = \int d^4 x \ \sqrt{- \det{(g_{\mu \nu})}} \ (\mathcal{L}_{\rm E-H} + \mathcal{L}_{\rm YM} + \alpha \mathcal{L}_4^1) \,, \label{model}
\end{equation}
where $g_{\mu \nu}$ is the metric tensor, $\mathcal{L}_{\rm E-H}$ is the Einstein-Hilbert Lagrangian,
\begin{equation}
\mathcal{L}_{\rm YM} \equiv - \frac{1}{4} F_{\mu \nu}^a F^{\mu \nu}_a \,,
\end{equation}
is the canonical kinetic term of $A_\mu$, and
\begin{equation}
F_{\mu \nu}^a \equiv \nabla_\mu A_\nu^a - \nabla_\nu A_\mu^a + g \epsilon^a_{\; \; bc} A_\mu^b A_\nu^c \,,
\end{equation}
where $g$ is the coupling constant of the group whereas the group structure constants are given by the Levi-Civita symbol $\epsilon_{abc}$.

{\it The autonomous dynamical system} - In order to sustain a homogeneous and isotropic background, the cosmic triad is described by
\begin{equation}
A_\mu^a = a \psi \delta_\mu^a \,,
\end{equation}
where $a \psi$ represents the homogeneous norm of the triad and $a$ is the scale factor in the Friedmann-Lemaitre-Robertson-Walker spacetime.  In terms of the dimensionless quantities
\begin{eqnarray}
x &\equiv& \frac{\dot{\psi}}{\sqrt{2} m_P H} \,, \nonumber \\
y &\equiv& \frac{\psi}{\sqrt{2} m_P} \,, \nonumber \\
z &\equiv& \frac{\psi}{\sqrt{2 m_P H}} \,, \label{def}
\end{eqnarray}
where $m_P$ is the reduced Planck mass, $H$ is the Hubble parameter, and a dot represents a derivative with respect to the cosmic time $t$, the field equations coming from the terms proportional to $\delta g_{\mu \nu}$ and $\delta A_\mu$ in $\delta S = 0$, with $S$ as in Eq. (\ref{model}), turn out to be
\begin{eqnarray}
&& y^2 + 2xy + x^2 + 2g^2z^4 + \alpha (-32 y^4 - 188 xy^3 + 10 x^2 y^2) \nonumber \\
&& = 1 \,, \label{Eequation1}
\end{eqnarray}
\begin{eqnarray}
&& y^2 + 2xy +x^2 + 2g^2z^4 \nonumber \\
&& + \alpha(-340 y^4 + 124 y^4 \epsilon + 316 xy^3 + 614 x^2 y^2 + 104 \sqrt{2} y^3 p) \nonumber \\
&& = -3 + 2\epsilon \,, \label{Eequation2}
\end{eqnarray}
\begin{eqnarray}
&& \frac{p}{\sqrt{2}} + 2y + 3x - y\epsilon + 4g^2 \frac{z^4}{y} \nonumber \\
&& + \alpha(-218 y^3 + 30 xy^2 + 94 y^3 \epsilon + 10 x^2 y + 5\sqrt{2} y^2 p) \nonumber \\
&& = 0 \,, \label{Aequation}
\end{eqnarray}
where
\begin{equation}
p \equiv \frac{\ddot{\psi}}{m_P H^2} \,,
\end{equation}
is a dimensionless quantity and
\begin{equation}
\epsilon \equiv - \frac{\dot{H}}{H^2} \,,
\end{equation}
is one of the standard slow-roll parameters.
Eq. (\ref{Aequation}) is redundant, it being already included in the Einstein field equations (\ref{Eequation1}) and (\ref{Eequation2}).  However, in a dynamical systems approach, Eq. (\ref{Eequation1}) acts as a constraint for the dimensionless parameters $x,y,z$ whose evolution equations are
\begin{eqnarray}
x' &=& \frac{p}{\sqrt{2}} + \epsilon x \,, \label{asymx} \\
y' &=& x \,, \label{asymy} \\
z' &=& z \left(\frac{x}{y} + \frac{\epsilon}{2}\right) \,, \label{asymz}
\end{eqnarray}
whereas Eqs. (\ref{Eequation2}) and (\ref{Aequation}) serve as a way to solve both $p$ and $\epsilon$ in terms of $x,y,z$.  In the previous expressions, a prime represents a derivative with respect to the e-folds number $N \equiv \int H dt$.

{\it Asymptotic behaviour and dark energy} - This autonomous dynamical system enjoys a nice asymptotic behaviour that leads to the description of two coexistent but artificial radiation perfect fluids, one associated to the Yang-Mills Lagrangian with both positive energy density and pressure which we will call the positive fluid, and the other associated to the $S$ term with both negative energy density and pressure which we will call the negative fluid.  Such asymptotic behaviour is given by 
\begin{eqnarray}
y &\rightarrow& \beta x \,, \nonumber \\
z &\rightarrow& \gamma x \,, \nonumber \\
x &\rightarrow& \infty \,, \nonumber \\
{\rm when} \;\; N &\rightarrow& \infty \,,  \label{asymlimit}
\end{eqnarray}
with $\beta, \gamma \in \mathbb{R}$.  Indeed, Eq. (\ref{asymy}) is consistent with this behaviour as $x' - \beta^{-1} x = 0$, i.e., $x \propto e^{N/\beta} \rightarrow \infty$.  As can be checked, Eqs. (\ref{Eequation1}) and (\ref{asymx}) - (\ref{asymz}) are satisfied simultaneously in the asymptotic regime for
\begin{eqnarray}
\beta &=& \frac{29}{11} \,, \nonumber \\
\gamma &=& \frac{\sqrt{29} (42837 \alpha)^{1/4}}{11 \sqrt{|g|}} \,, \label{asymbe}
\end{eqnarray}
which, in turn, makes $\epsilon \rightarrow 0$ (and, therefore, $\omega \rightarrow -1$).  
\begin{figure}
\subfloat[\label{y/xvsN}]{%
  \includegraphics[height=3.5cm,width=.48\linewidth]{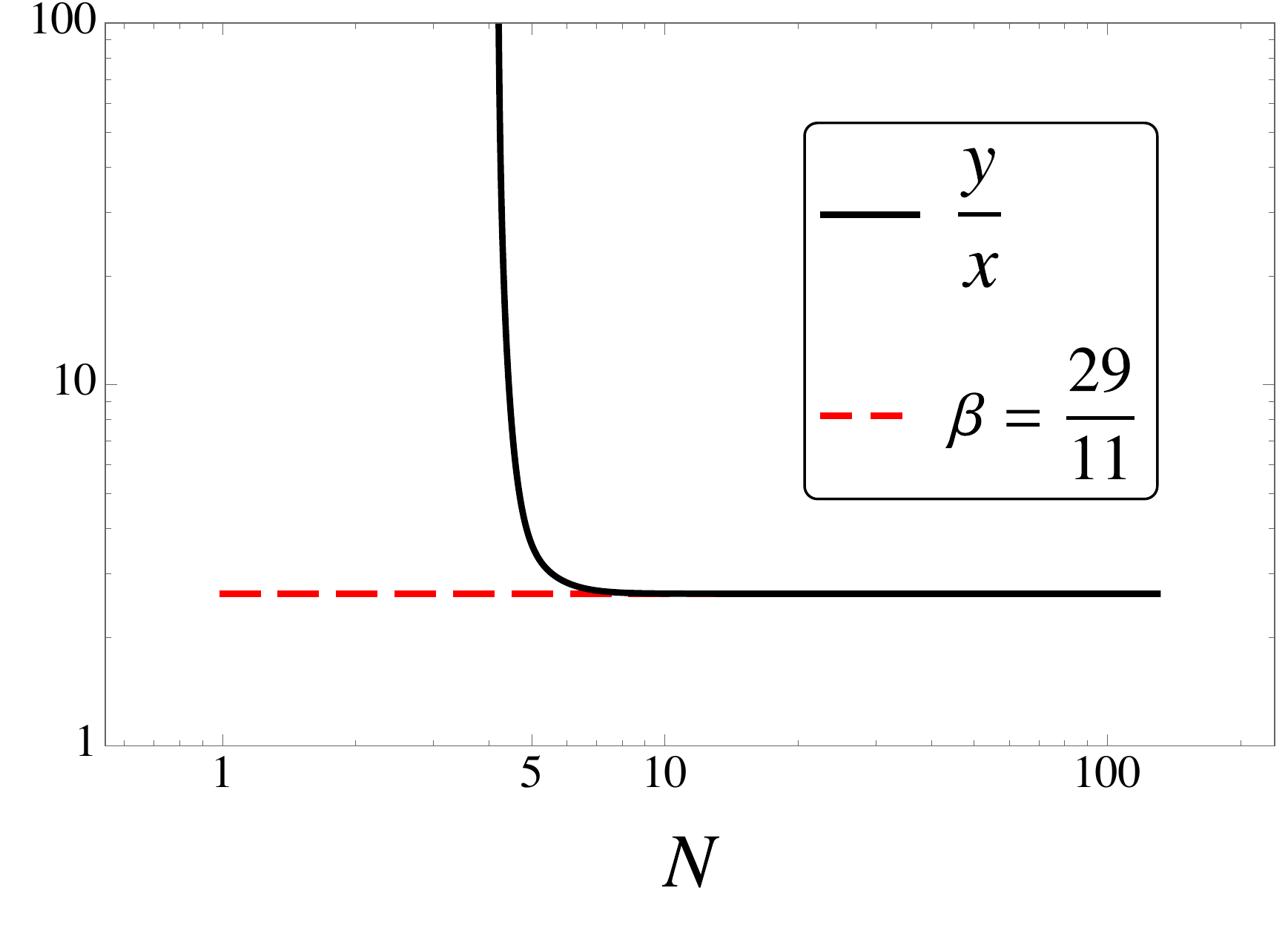}%
}\hfill
\subfloat[\label{z/xvsN}]{%
  \includegraphics[height=3.5cm,width=.48\linewidth]{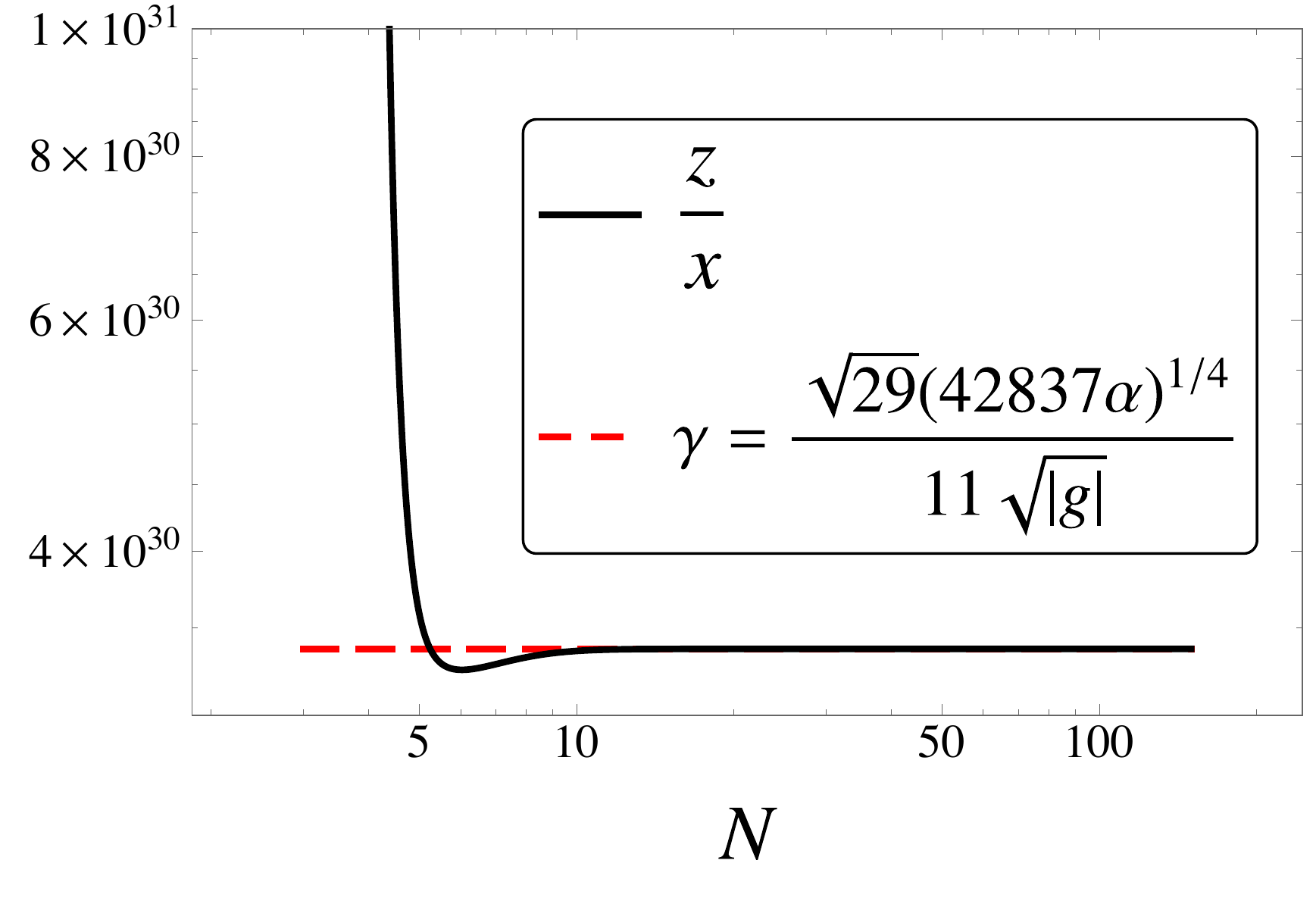}%
}\\
\subfloat[\label{rhoa-rhoymvsN}]{%
  \includegraphics[height=3.5cm,width=.48\linewidth]{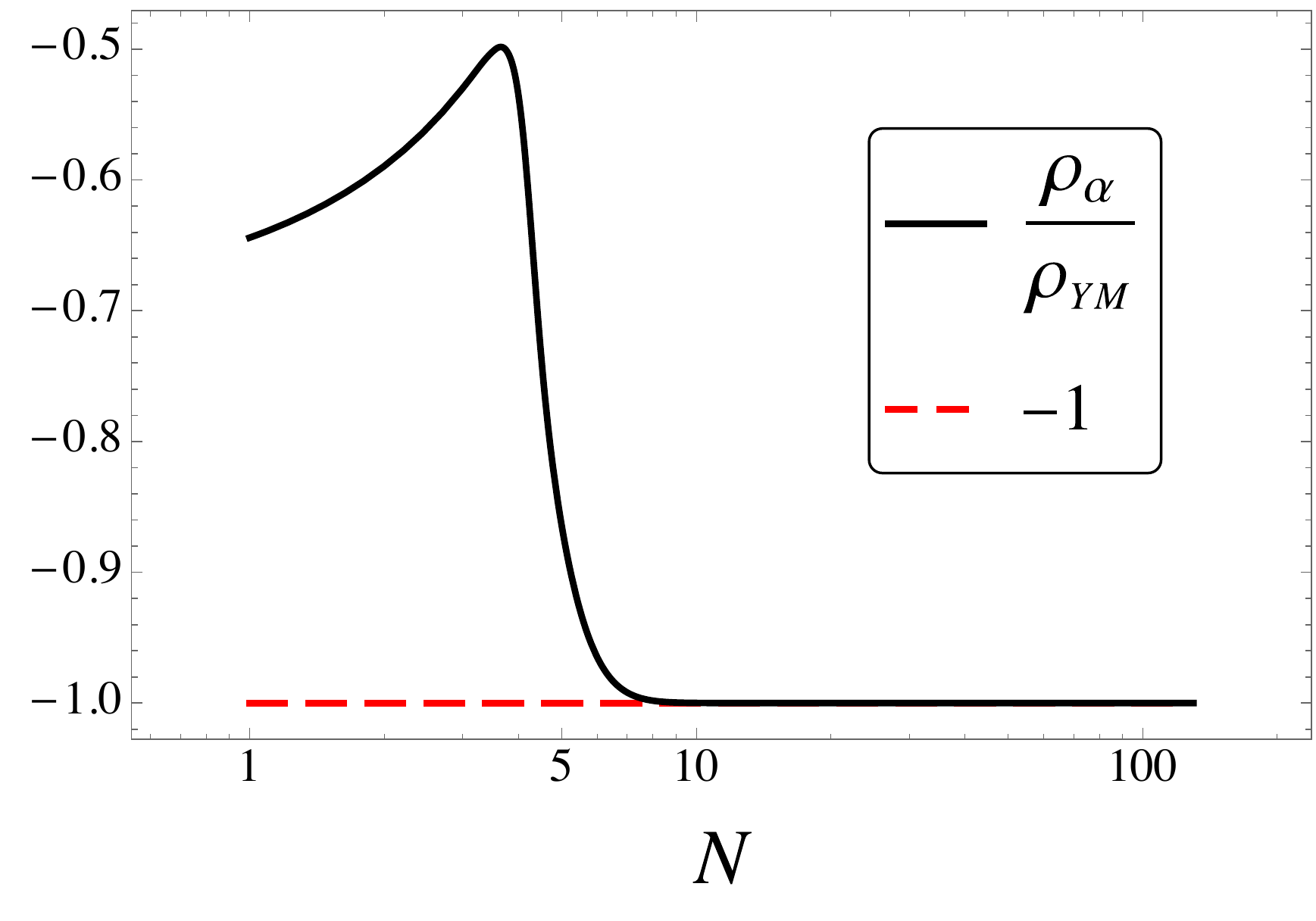}%
}\hfill
\subfloat[\label{rhoym-rhoavsN}]{%
  \includegraphics[height=3.5cm,width=.48\linewidth]{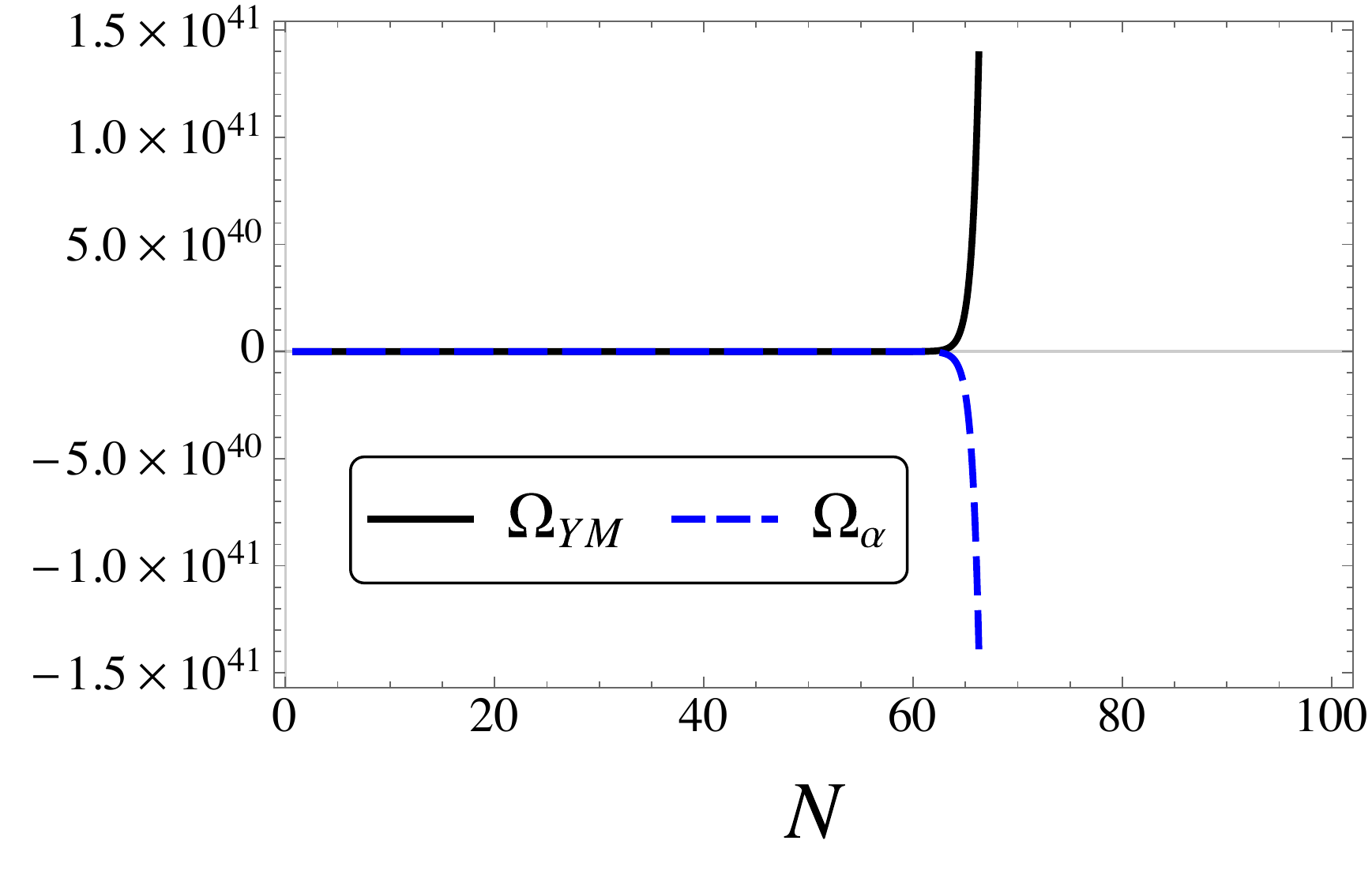}%
}\\
\subfloat[\label{HvsN}]{%
  \includegraphics[height=3.5cm,width=.48\linewidth]{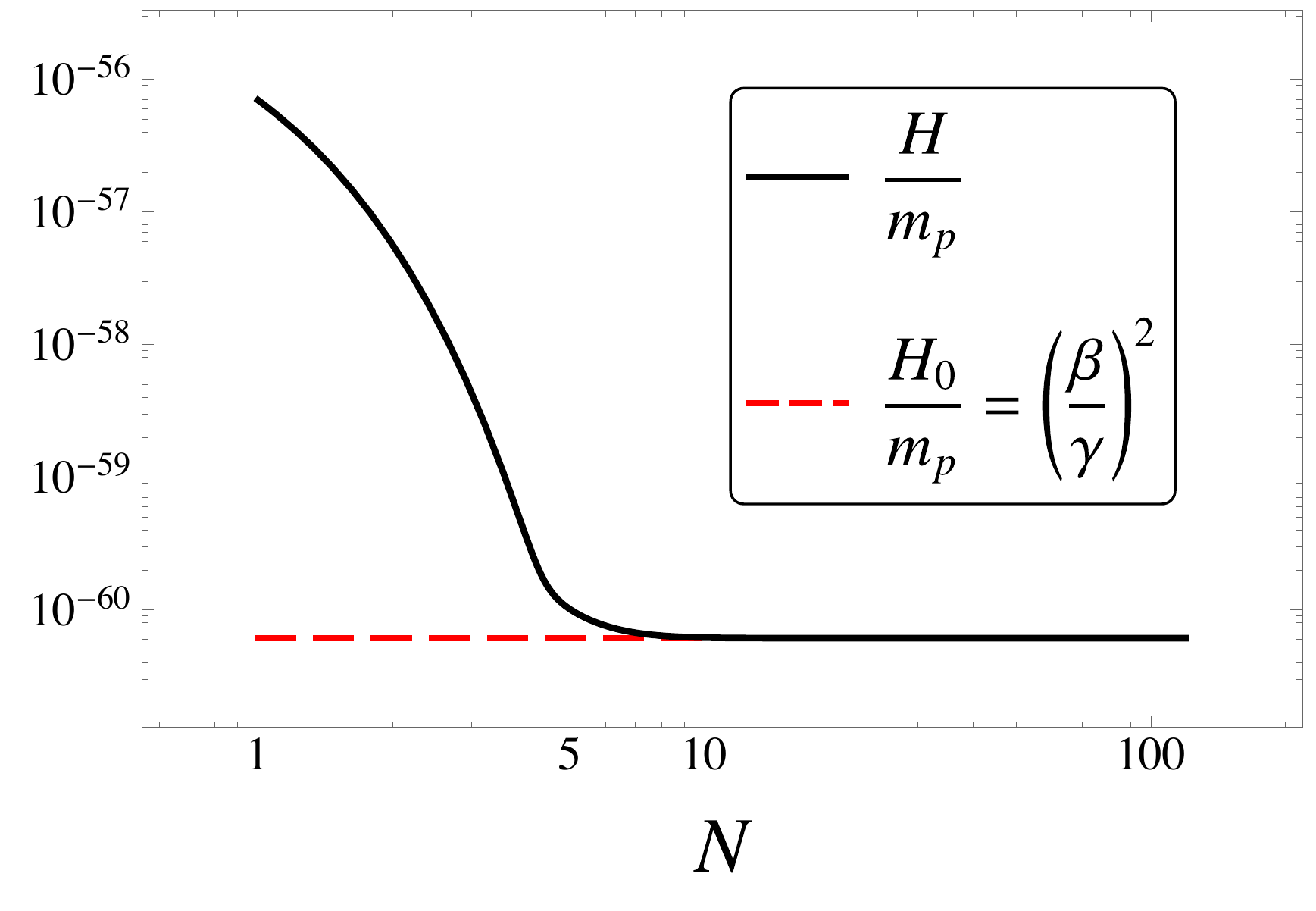}%
}\hfill
\subfloat[\label{wavsN}]{%
  \includegraphics[height=3.5cm,width=.48\linewidth]{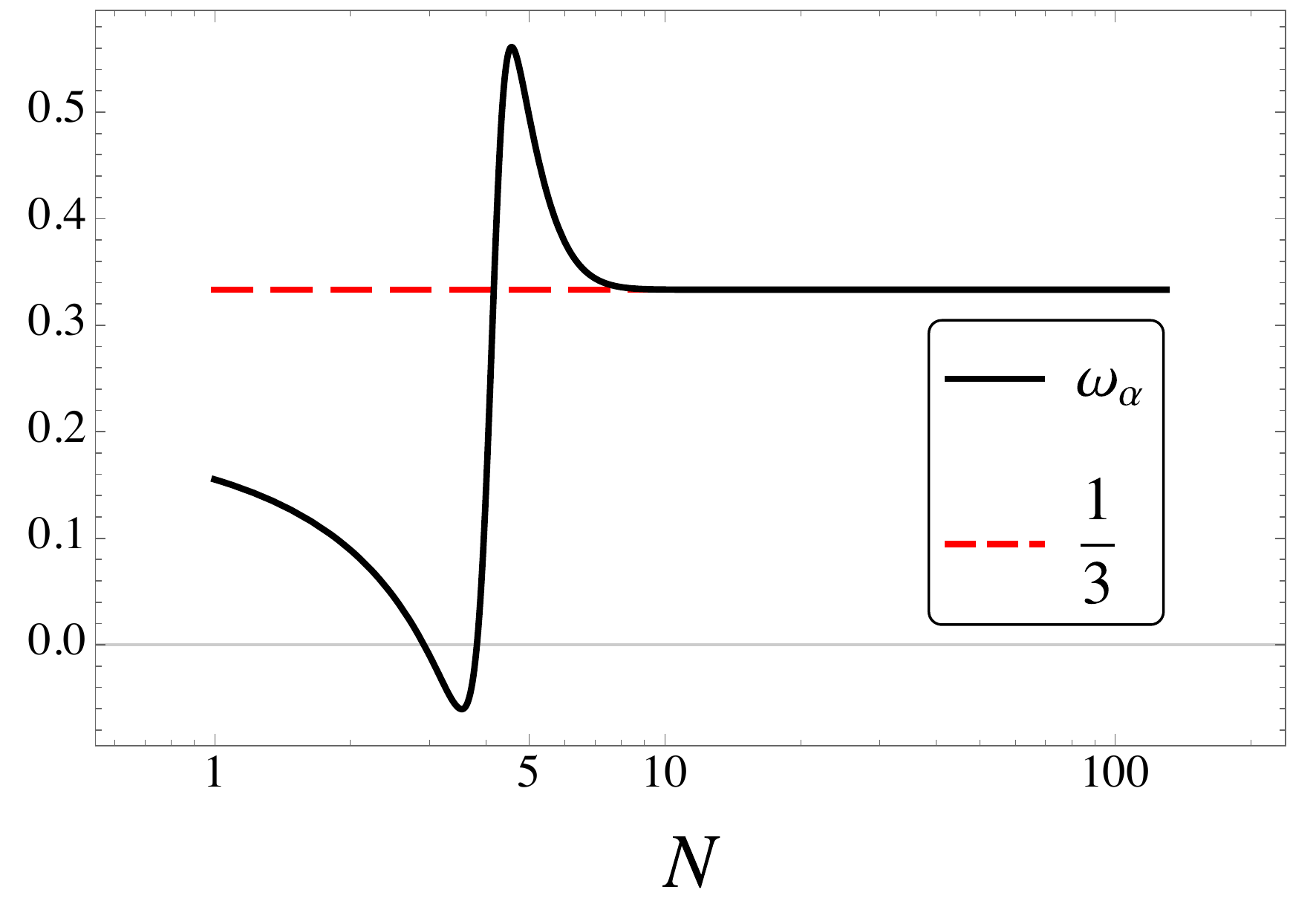}%
}\\
\subfloat[\label{wtotvsN}]{%
  \includegraphics[height=3.5cm,width=.48\linewidth]{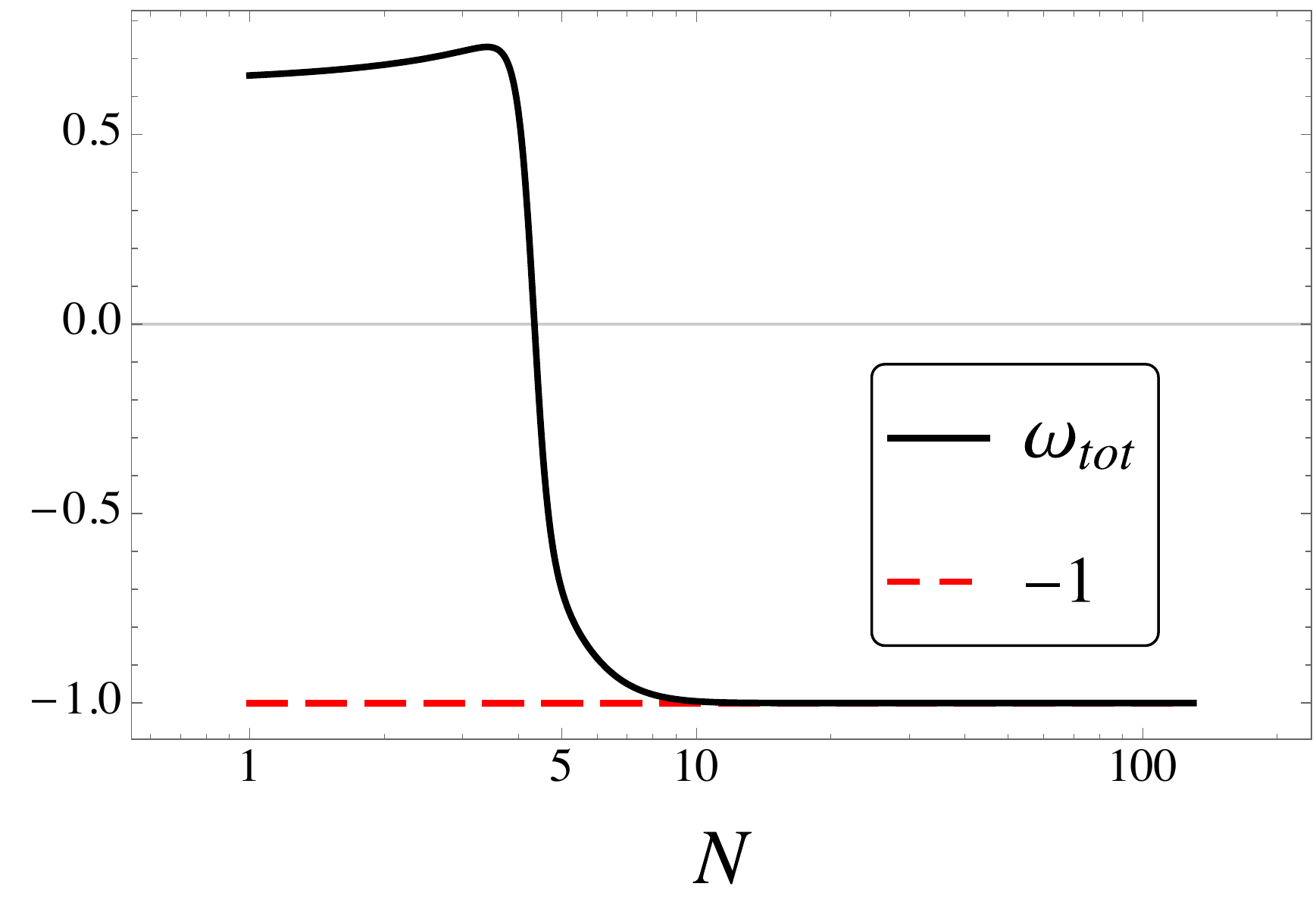}%
}    
\caption{Dark energy - Numerical solutions and asymptotes for $\alpha = 0.01$, $g = -4.37 \times 10^{-61}$, and initial conditions $x_0 = -0.187568$, $y_0 = 2$, and $z_0 = 7 \times 10^{27}$. a) This figure shows the numerical solution for $y/x$ vs. $N$ (black continuous curve) as well as the asymptotic behaviour $y/x = \beta$ (red dashed curve) with $\beta$ as in Eq. (\ref{asymbe}); the pair of curves overlap from $N \approx 10$ onwards.  b) Same as before but for $z/x$.  c) This figure confirms that $\rho_\alpha \rightarrow - \rho_{\rm YM}$ in the asymptotic limit. d) This figure presents $\Omega_{\rm YM}$ vs. $N$ (black continuous curve) and $\Omega_\alpha$ vs. $N$ (blue dashed curve); we can see that the absolute values of the two density parameters grow exponentially with $N$. e) This figure presents $H/m_P$ vs. $N$; although $\rho_{\rm YM}$ and $|\rho_\alpha|$ grow exponentially, $\rho_{\rm tot} \propto \sqrt{H}$ approaches asymptotically a finite constant value. f) This figure shows $\omega_\alpha$ vs. $N$; in the asymptotic limit, the negative fluid behaves as radiation. g) This figure shows $\omega_{\rm tot}$ vs. $N$; in the asymptotic limit, the total fluid behaves as a cosmological constant.}
\label{asymfig1}
\end{figure}
Negative values for $\alpha$ would render $\gamma \in \mathbb{C}$ disabling the asymptotic behaviour; in contrast, positive and negative values for $g$ are allowed.
Figs. \ref{y/xvsN} and \ref{z/xvsN} show the asymptotic behaviour for $y/x$ and $z/x$ for a chosen set of initial conditions. As stated before, the system exhibits an asymptotic dynamical fine-tuning mechanism, as the absolute values of the energy densities of the negative fluid ($|\rho_\alpha|$) and of the positive fluid ($\rho_{\rm YM}$) grow exponentially but, nevertheless, matches almost precisely,  irrespective of the initial conditions\footnote{Unless they are near enough an attractor of the dynamical system. \label{f2}},  so that $\rho_{\rm tot} = \rho_{\rm YM} + \rho_\alpha$ approaches a finite constant value.  To see this, we can easily extract $\Omega_\alpha \equiv \rho_\alpha / \rho_{\rm tot}$ and $\Omega_{\rm YM} \equiv \rho_{\rm YM} / \rho_{\rm tot}$ from Eq. (\ref{Eequation1}):
\begin{eqnarray}
\Omega_{\rm YM} &=& y^2 + 2 x y + x^2 + 2 g^2 z^4 \,, \nonumber \\
\Omega_\alpha &=& \alpha (-32 y^4 - 188 x y^3 + 10 x^2 y^2) \,,
\end{eqnarray}
and check that in the asymptotic limit given by Eqs. (\ref{asymlimit}) and (\ref{asymbe}), $\rho_\alpha \rightarrow - \rho_{\rm YM}$.
This is confirmed by the numerical solution presented in Fig. \ref{rhoa-rhoymvsN}.
Fig. \ref{rhoym-rhoavsN} shows the exponential growth with $N$ of both $\Omega_{\rm YM}$ and $|\Omega_\alpha|$ whereas Fig. \ref{HvsN} reveals the predicted behaviour for $\rho_{\rm  tot} \propto \sqrt{H}$.  As observed in the figures, the apparent breakdown of the classical regime due to the exponential growth of the absolute values of the energy densities is disproved by the good behaviour of $H < m_P$.  The equation of state parameter for each fluid can also be studied by extracting the pressures $P_{\rm YM}$ and $P_\alpha$ from Eqs. (\ref{Eequation1}) and (\ref{Eequation2}):
\begin{eqnarray}
P_{\rm YM} &=& \frac{\rho_{\rm tot}}{3}(y^2 + 2 x y + x^2 + 2 g^2 z^4) \,, \nonumber \\
P_\alpha &=& \frac{\rho_{\rm tot}}{3}[\alpha (-340 y^4 + 124 y^4 \epsilon + 316 x y^3 \nonumber \\
&& + 614 x^2 y^2 + 104 \sqrt{2} y^3 p)] \,.
\end{eqnarray}
Of course, $\omega_{\rm YM} = 1/3$, whereas $\omega_\alpha$ is a complicated function of $N$ that goes, in the asymptotic limit given by Eqs. (\ref{asymlimit}) and (\ref{asymbe}), to $1/3$;  this is numerically confirmed in Fig. \ref{wavsN}.  Regarding the equation of state parameter for the whole fluid, we have asymptotically $\omega_{\rm tot} \rightarrow -1$;  this is analytically the case as shown just below Eq. (\ref{asymbe}) and it is numerically confirmed in Fig. \ref{wtotvsN}.
Such an interesting built-in self-tuning mechanism to generate an equation of state parameter $\omega \simeq -1$, in agreement with the observed equation of state parameter for dark energy $\omega =  -1.006 \pm 0.045$ \cite{Ade:2015xua,Ade:2015rim}, could not be an ideal candidate to explain the dark energy if the asymptotic value for $H$ did not correspond to the observed value today $H_0 = 9.03 h \times 10^{-61} m_P$ with $h = 0.678 \pm 0.009$ \cite{Ade:2015xua}.  From the definitions in Eq. (\ref{def}) and the asymptotic behaviour described in Eqs. (\ref{asymlimit}) and (\ref{asymbe}), we obtain
\begin{equation}
\frac{H}{m_P} \rightarrow \left(\frac{\beta}{\gamma}\right)^2 = \frac{29 |g|}{\sqrt{42837 \alpha}} \,,
\end{equation}
which is consistent with the numerical solution for $H$ in Fig. \ref{HvsN}. This reveals that $H$ can reproduce its presently observed value for a low enough value of $|g|$: $|g| \lesssim 7 H_0/m_P \approx 4 \times 10^{-60}$, irrespective of the initial conditions (for $\alpha > 0$, see footnote \ref{f2}).  The insensitivity to the initial conditions can be observed comparing Figs. \ref{asymfig1} and \ref{asymfig2}.
\begin{figure}
\subfloat[\label{y/xvsN2}]{%
  \includegraphics[height=3.5cm,width=.48\linewidth]{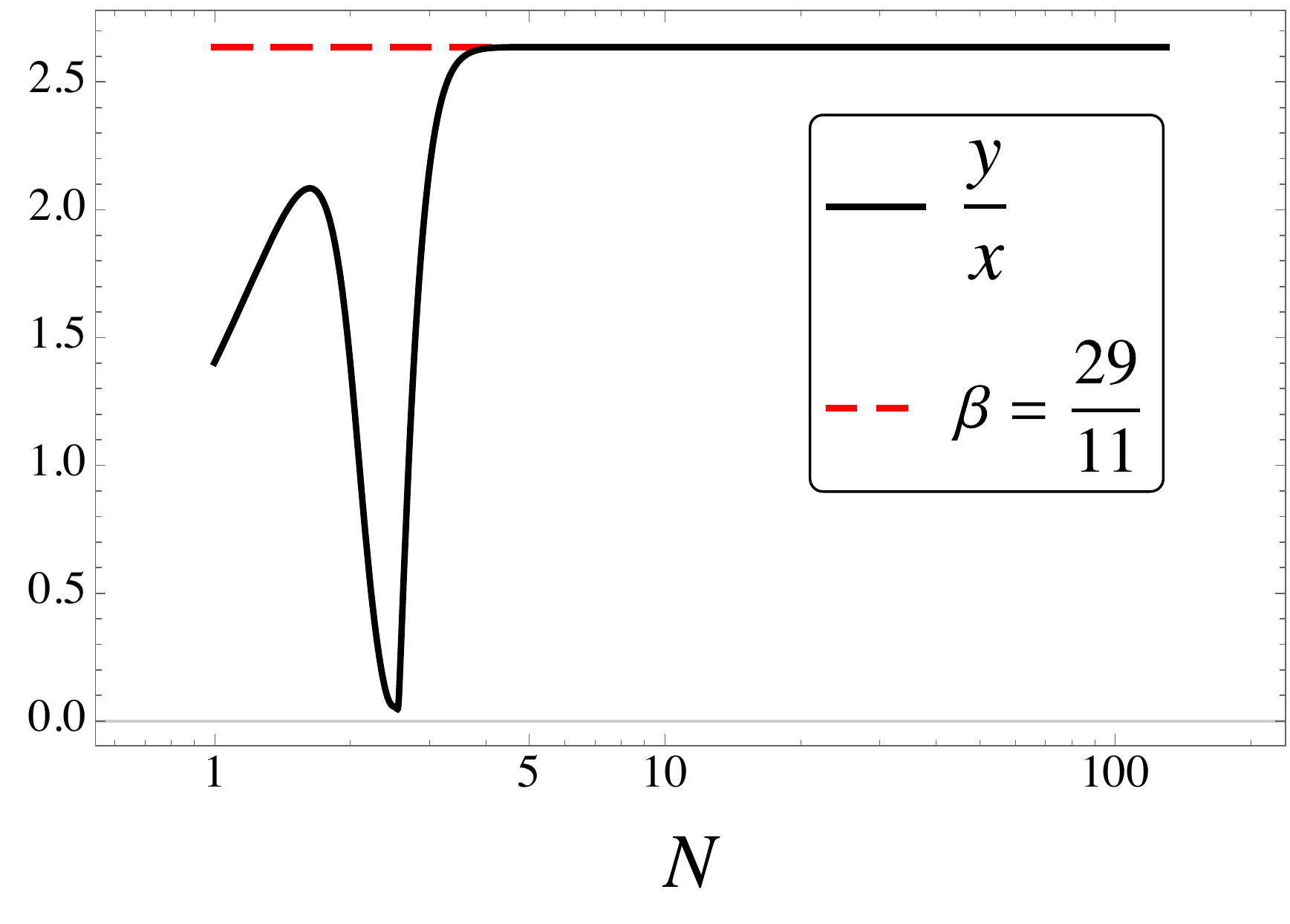}%
}\hfill
\subfloat[\label{z/xvsN2}]{%
  \includegraphics[height=3.5cm,width=.48\linewidth]{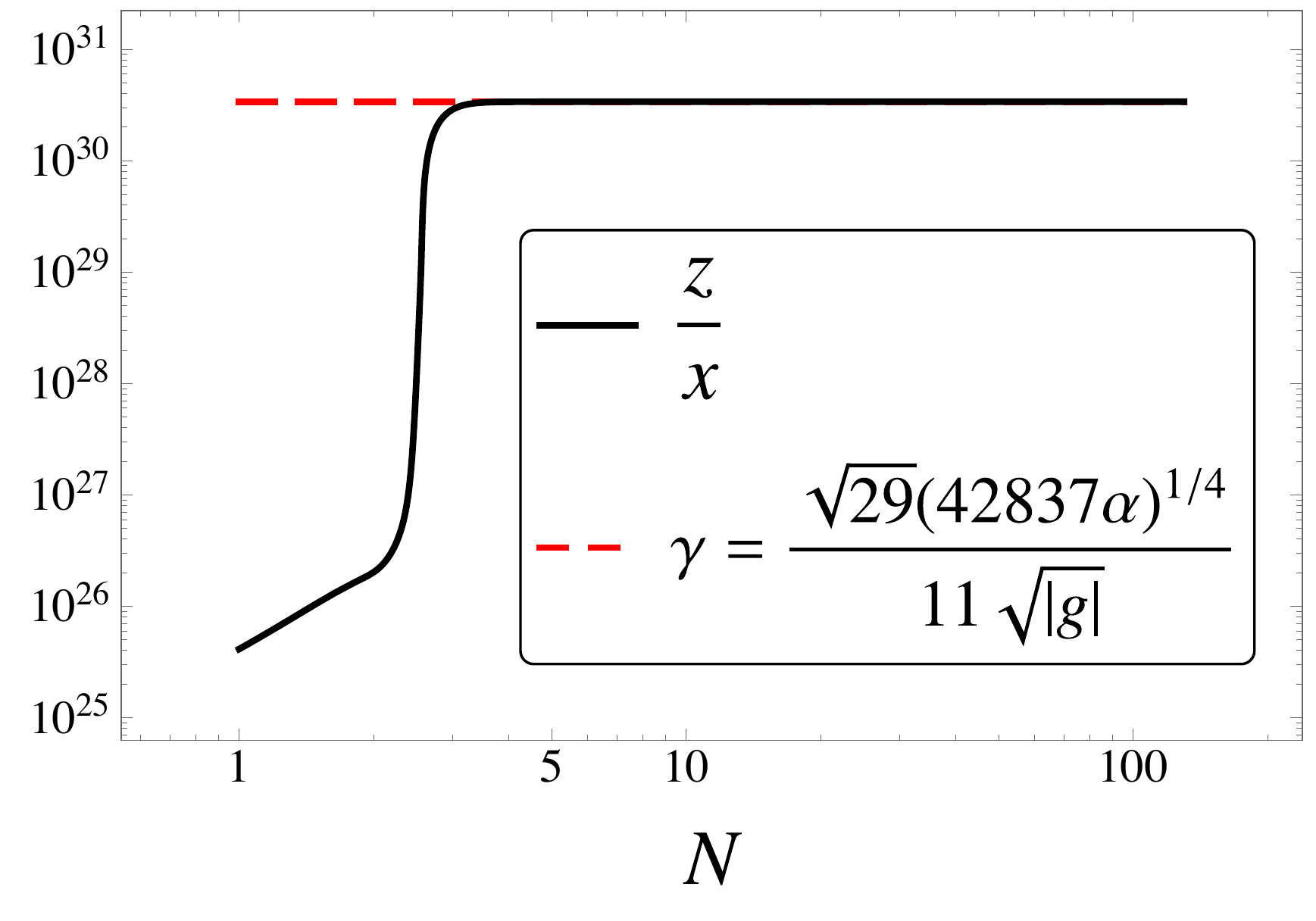}%
}\\
\subfloat[\label{rhoa-rhoymvsN2}]{%
  \includegraphics[height=3.5cm,width=.48\linewidth]{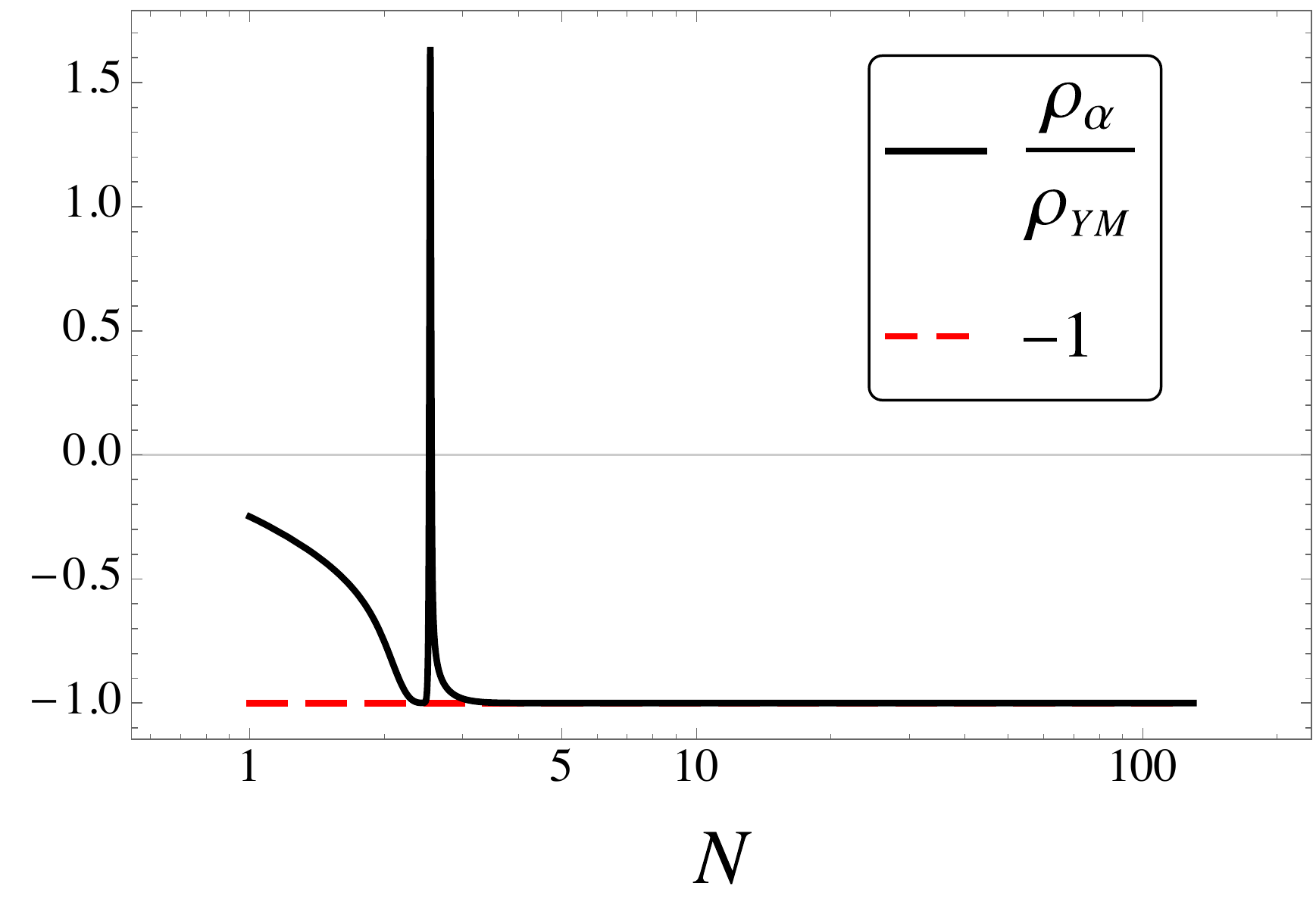}%
}\hfill
\subfloat[\label{rhoym-rhoavsN2}]{%
  \includegraphics[height=3.5cm,width=.48\linewidth]{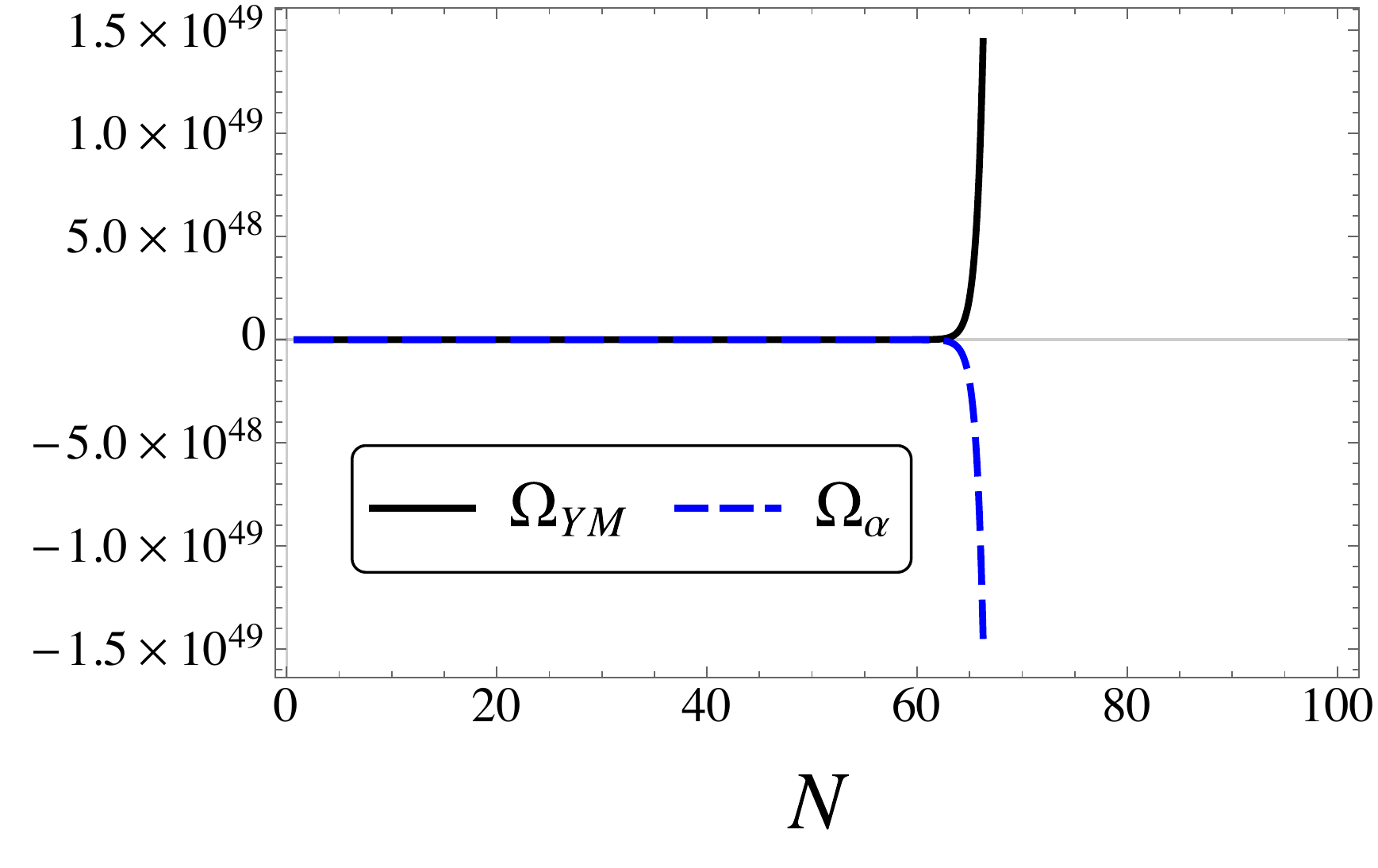}%
}\\
\subfloat[\label{HvsN2}]{%
  \includegraphics[height=3.5cm,width=.48\linewidth]{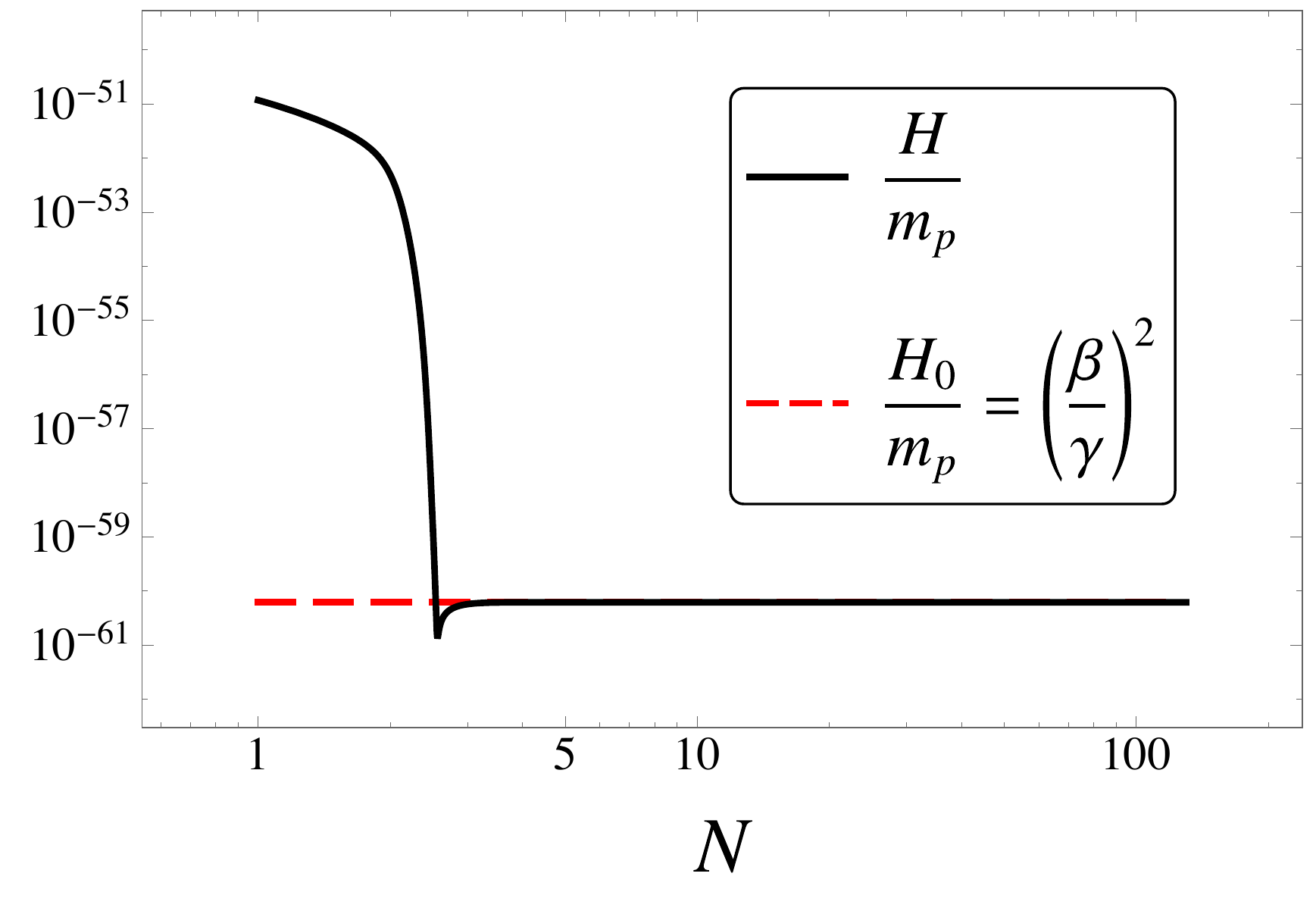}%
}\hfill
\subfloat[\label{wavsN2}]{%
  \includegraphics[height=3.5cm,width=.48\linewidth]{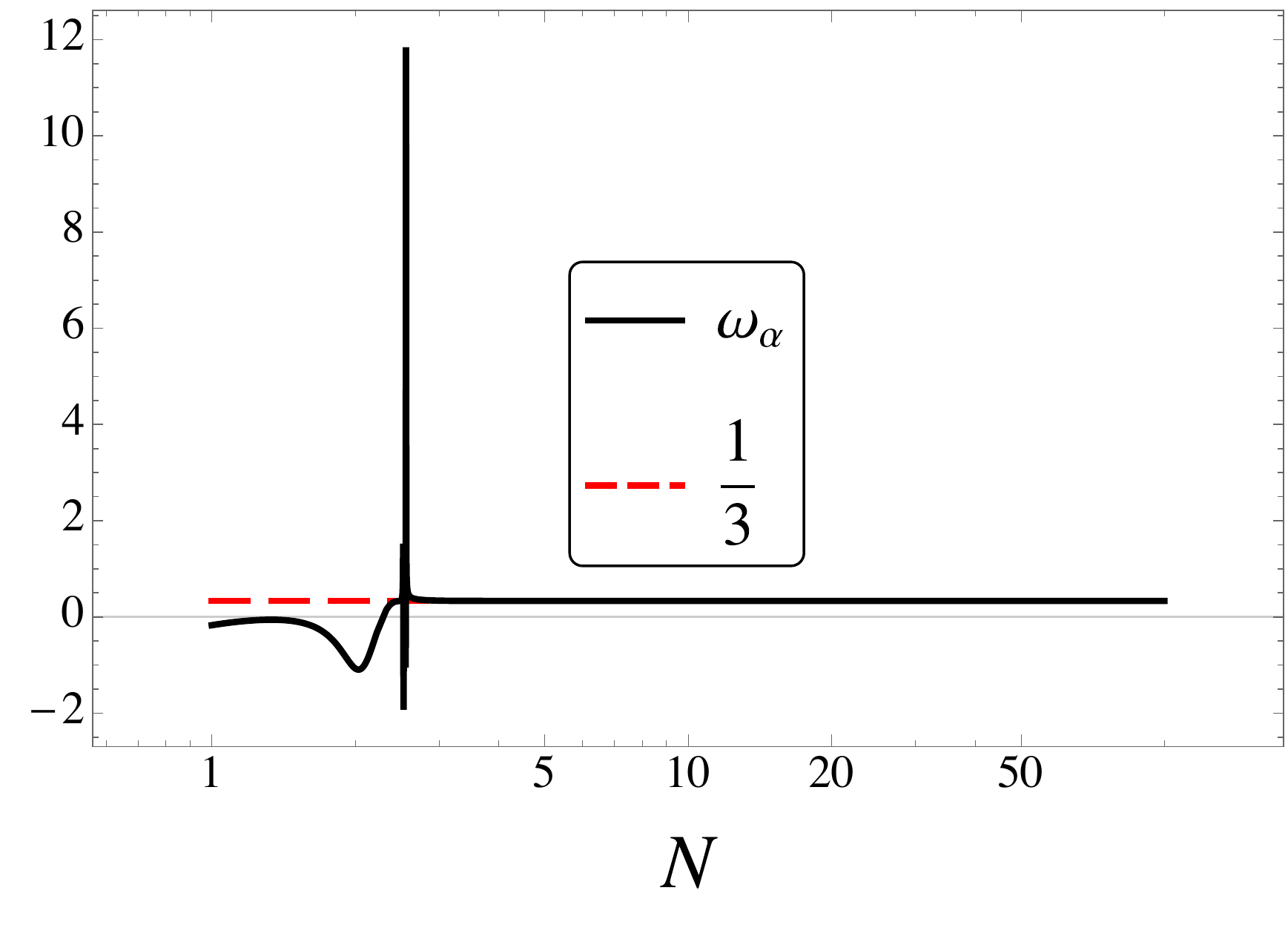}%
}\\
\subfloat[\label{wtotvsN2}]{%
  \includegraphics[height=3.5cm,width=.48\linewidth]{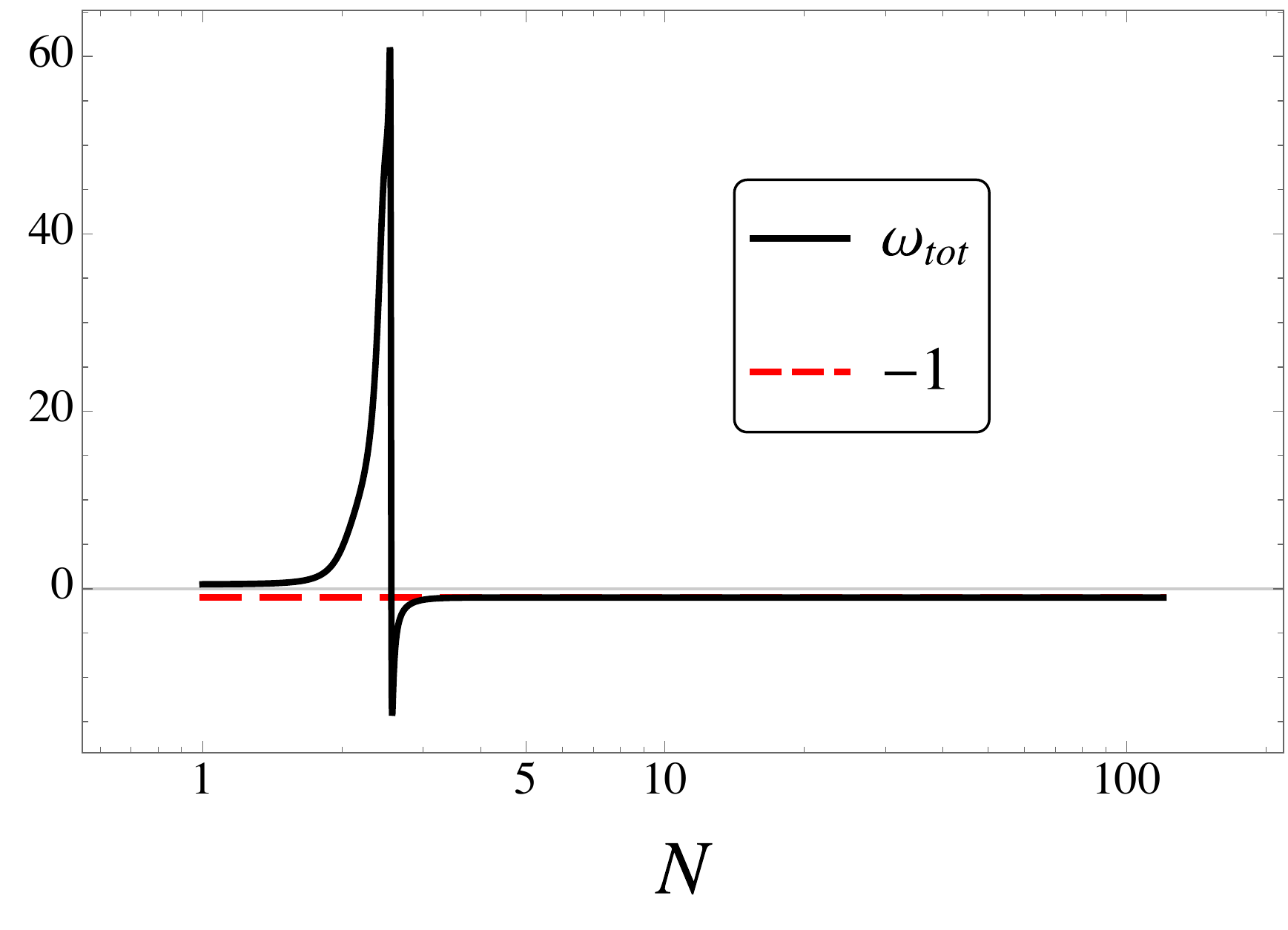}%
}    
\caption{Dark energy - Same as Fig. \ref{asymfig1} but with a different set of initial conditions:  $x_0 = 0.998999$, $y_0 = 10^{-3}$, and $z_0 = 10^{22}$.}
\label{asymfig2}
\end{figure}

\begin{figure}
\subfloat[\label{yvsN}]{%
  \includegraphics[height=3.5cm,width=.48\linewidth]{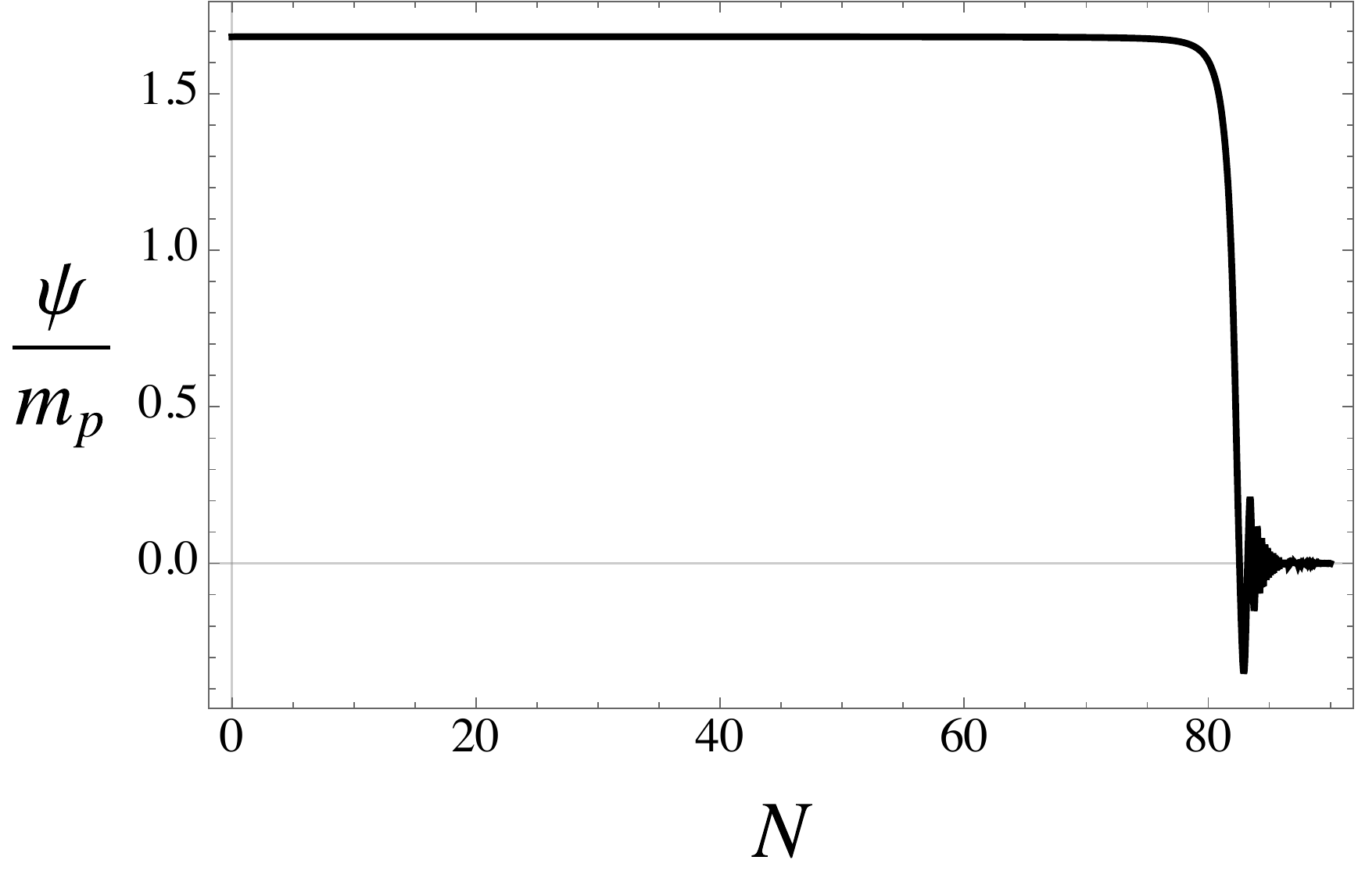}%
}\hfill
\subfloat[\label{phidotvsN}]{%
  \includegraphics[height=3.5cm,width=.48\linewidth]{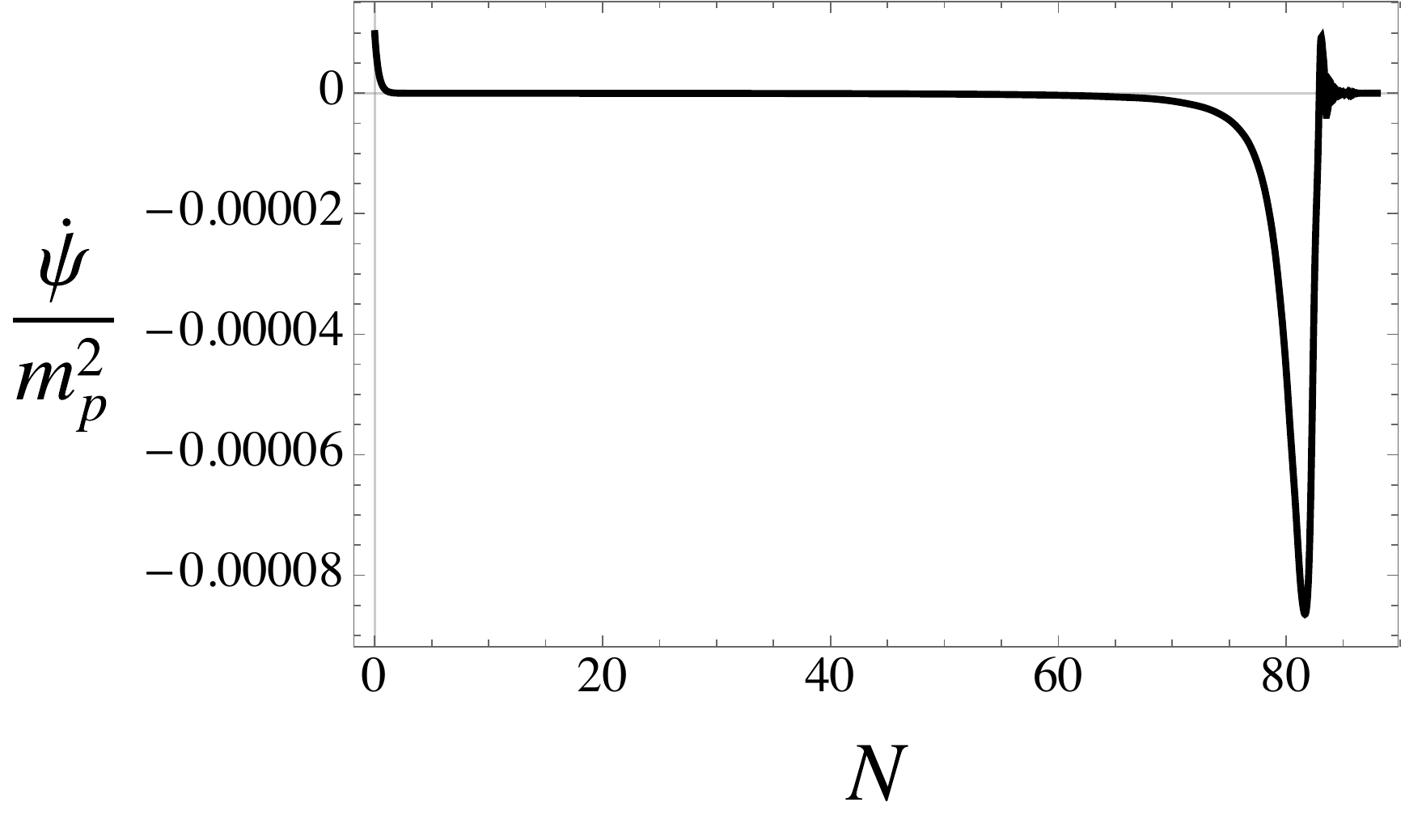}%
}\\
\subfloat[\label{epsilonetavsN}]{%
  \includegraphics[height=3.5cm,width=.48\linewidth]{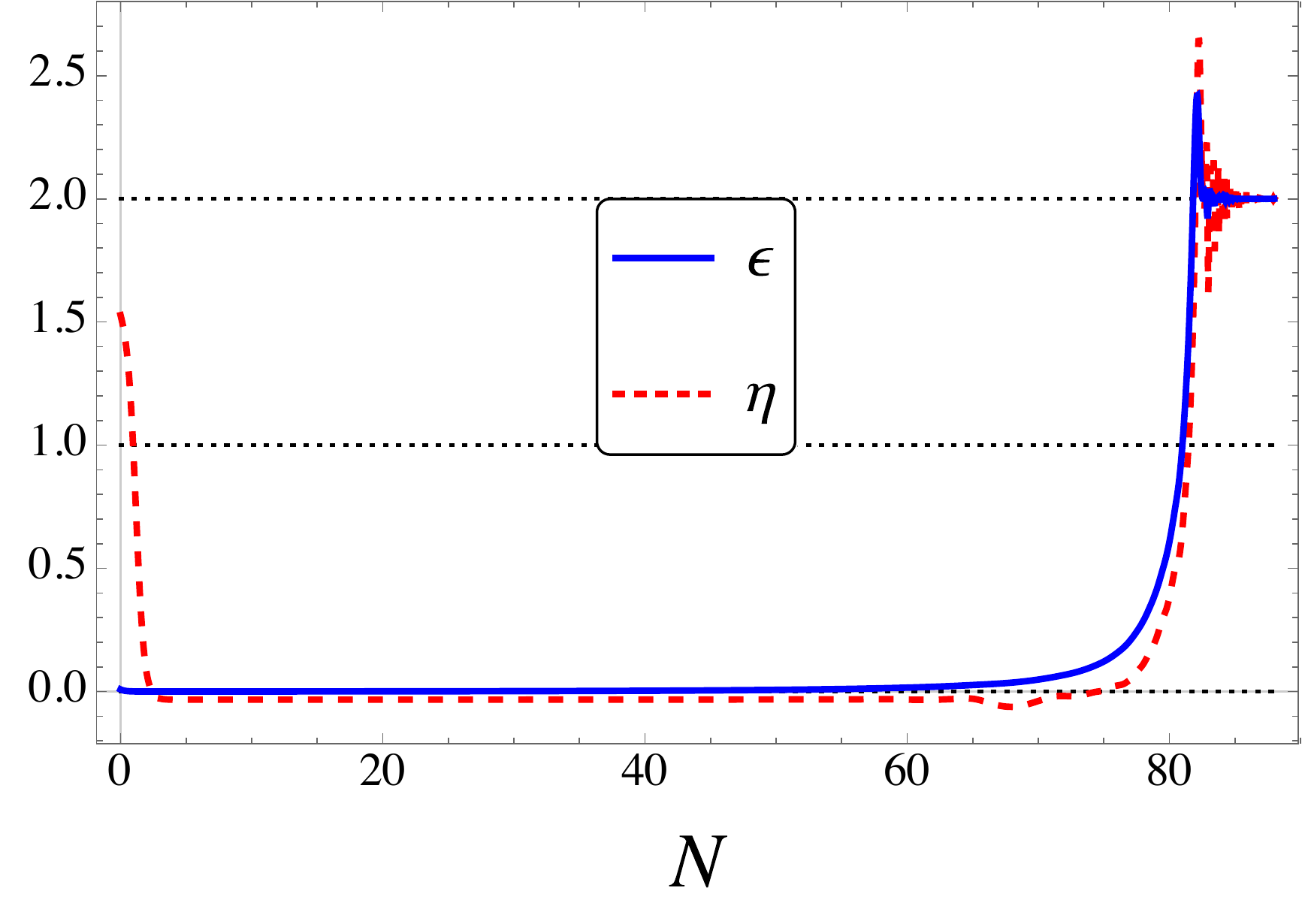}%
}\hfill
\subfloat[\label{17HvsN}]{%
  \includegraphics[height=3.5cm,width=.48\linewidth]{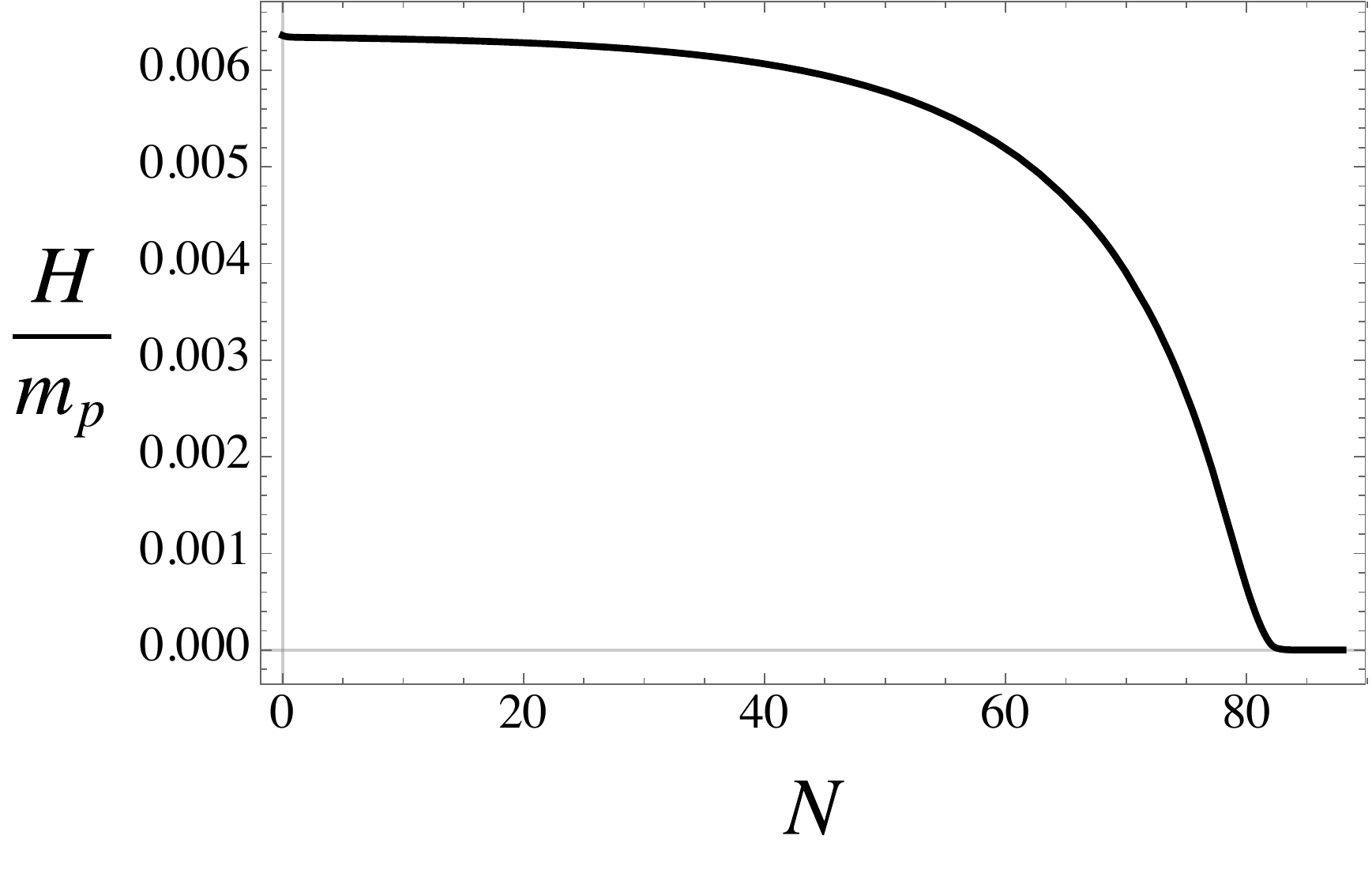}%
}\\
\subfloat[\label{rhoarhototvsN}]{%
  \includegraphics[height=3.5cm,width=.48\linewidth]{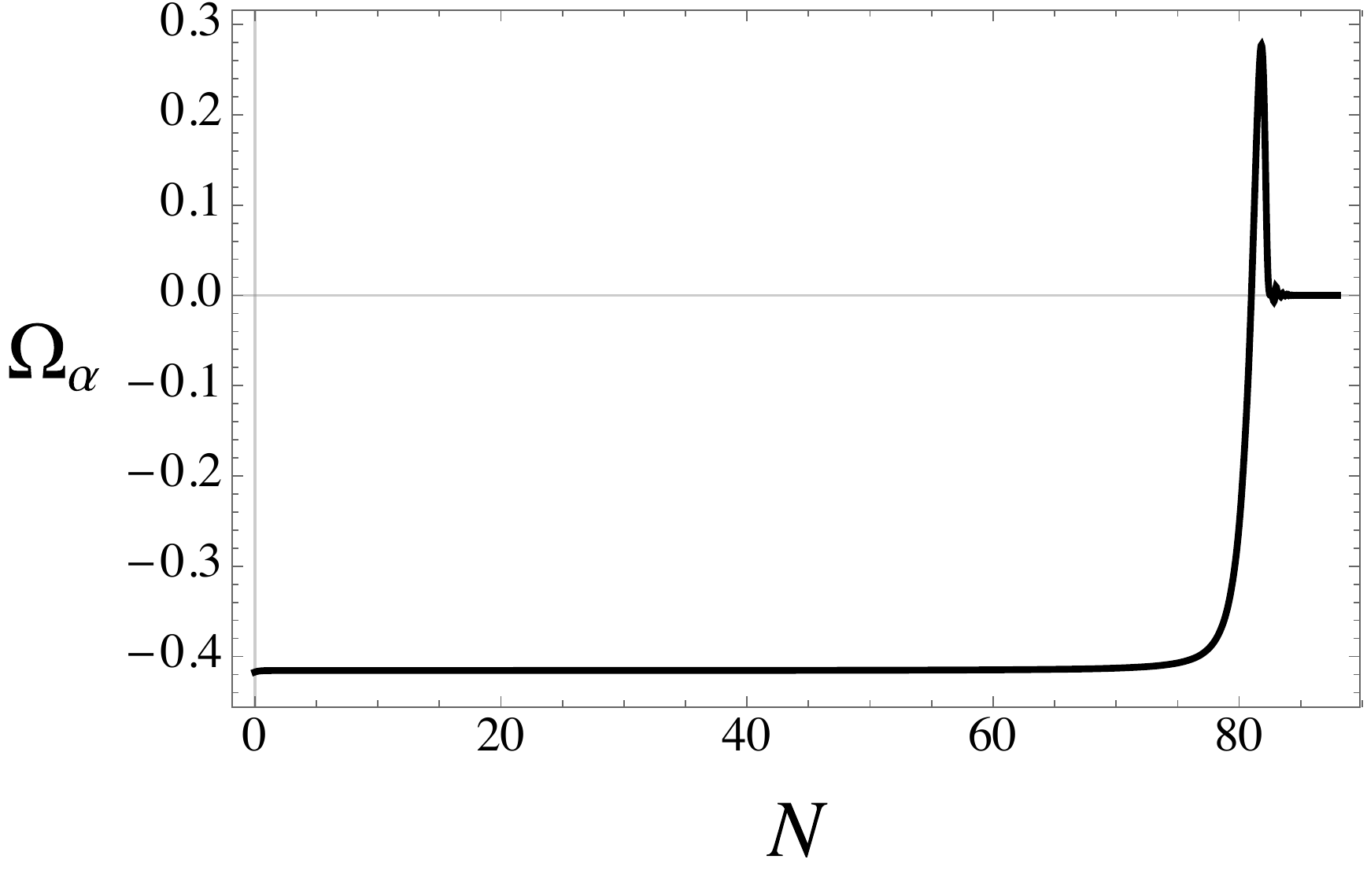}%
}\hfill
\subfloat[\label{PaPymvsN}]{%
  \includegraphics[height=3.5cm,width=.48\linewidth]{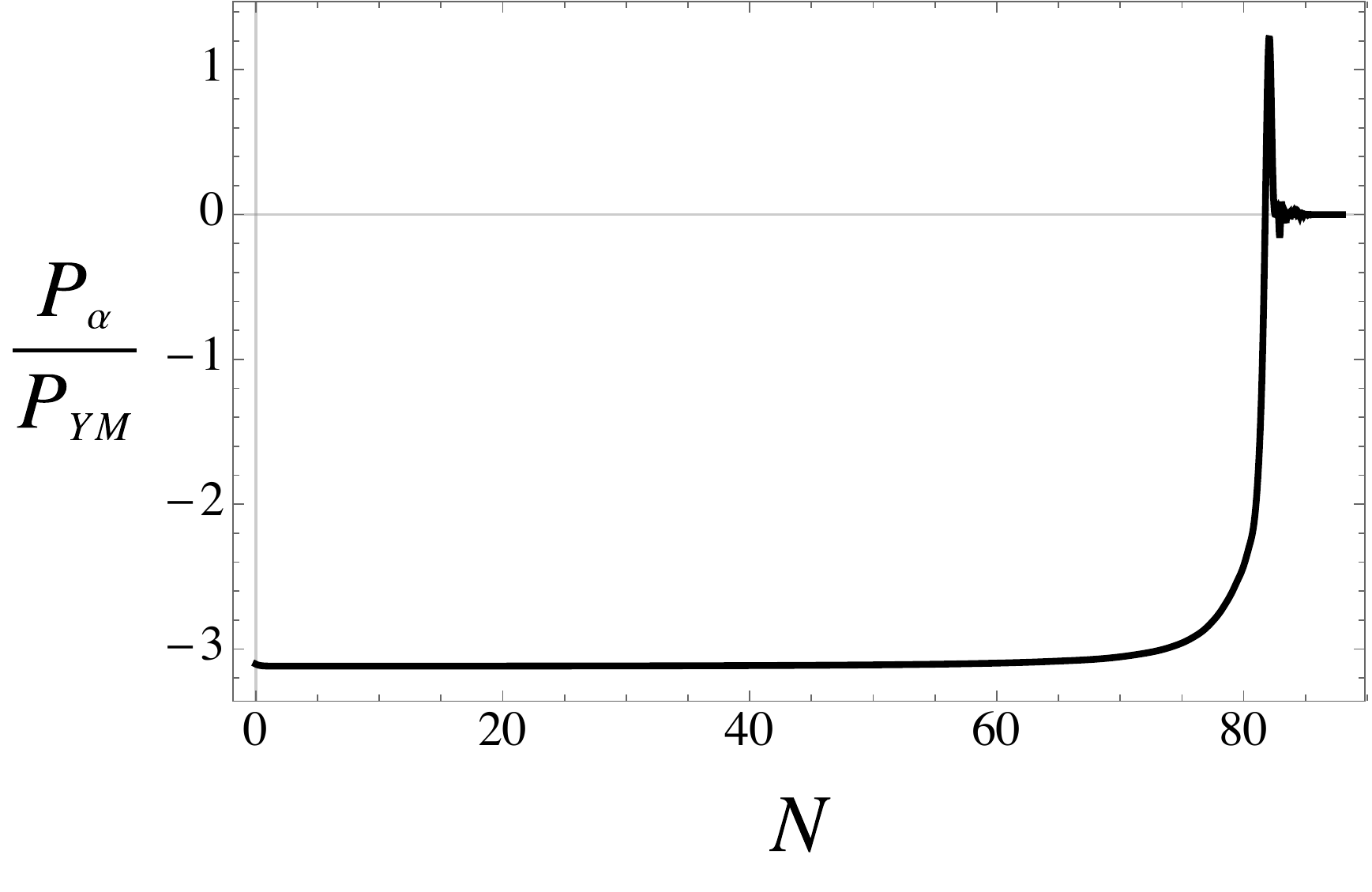}%
}
\caption{Primordial inflation - Numerical solutions for $\alpha = 0.00649$, $g = -0.0001$, and initial conditions $x_0 = 0.0011$, $y_0 = 1.189$, and $z_0 = 14.911$. a) This figure exhibits the slow-roll behaviour of $\psi/m_P$ during some 70 efolds and a final stage of damped oscillations signalling the radiation dominated epoch. b) Similar behaviour as in the previous figure but for  $\dot{\psi} / m_P^2$. c) This figure presents the slow-roll parameters $\epsilon$ (blue continuous curve) and $\eta$ (dashed red curve) vs. $N$; we can see that both $\epsilon$ and $|\eta|$ are well below 1 during some 70 efolds, implying a period of slow-roll inflation long enough to solve the classical problems of the standard cosmology; at the end of this stage, $\epsilon$ and $\eta$ go to 2, signalling the radiation dominated epoch. d) This figure presents $H/m_P$ vs. $N$; the energy scale of inflation is not in conflict with our classical treatment for low enough values of $|g|$. e) This figure shows $\Omega_\alpha$ vs. $N$; the $S$ term during inflation contributes with a negative energy density which is neither dominating nor negligible; it, nevertheless, becomes negligible by the end of this period. f) This figure shows $P_\alpha/P_{\rm YM}$ vs. $N$; the $S$ term during inflation contributes with a dominating negative pressure that becomes negligible by the end of this period letting the Yang-Mills term dominate and produce the radiation dominated epoch.}
\label{figinflation}
\end{figure}

{\it Primordial inflation} - Aside the asymptotic behaviour, the critical points of the dynamical system in Eqs. (\ref{asymx}) - (\ref{asymz}) have also been studied.  There exist 112 critical points, 51 of them being unrealistic as they correspond either to complex values for $y$ or $z$ or to real values for the same variables but with different signs, other 38 lead to $\epsilon \geq 1$, other 6 lead to $\epsilon < 0$ and the other 17 lead to $0 \leq \epsilon < 1$.  The latter ones are of special interest because they represent non-phantom accelerated expansion which, eventually, could be long enough but of finite duration to be ideal candidates to explain the primordial inflation. 8 out of these 17 critical points are unrealistic too because the inflationary period is only given for $H \rightarrow \infty$.  From the remaining 9 critical points, 8 are uninteresting, most of them because inflation is very short (a few efolds).  So we are left with 1 critical point, a saddle point indeed, that satisfies all the properties to be identified with a primordial inflationary period as described in  Fig. \ref{figinflation}. Unfortunately, and in contrast to the dark energy scenario, the primordial inflation in this model is strongly sensitive to the initial conditions and coupling constant $\alpha$. 

{\it Further exploration of the model} - The model presented above is so attractive, plausible, and well founded, at least as dark energy is concerned, that deserves further exploration.  One of the first things to do is to investigate whether the Hamiltonian is actually bounded from below\footnote{As the requirement that the dynamical equations must be, at most, second order in space-time derivatives is a necessary but not sufficient condition to avoid the Ostrogradski's instability.}.  In addition, a complete study of the cosmological perturbations is required in order to establish the robustness of the model against Laplacian and ghost instabilities, to check the perturbative stability of the isotropic solution, and to calculate the sound speed $c_s$ which is a distinctive feature of any dark energy model \cite{Hu:1998tj}.  Another aspect to explore is the possible attractor nature of the triad configuration in a more general anisotropic background (as is done for the Gauge-flation model in Ref. \cite{Maleknejad:2011jr}).  However, what is an urgent and necessary matter to investigate is whether the addition of $\kappa \mathcal{L}_4^2$ can evade the apparently strong constraints coming from the detection of the gravitational wave signal GW170817 \cite{TheLIGOScientific:2017qsa} and its electromagnetic counterpart GRB 170817A \cite{GBM:2017lvd,Monitor:2017mdv}:  a preliminary analysis, following the lines of Ref. \cite{Hertzberg:2017abn}, suggest that the coupling with the Riemann tensor in Eq. (\ref{riemann}) does not modify the gravitational waves speed;  this suggestion is strengthened by the fact that $\mathcal{L}_6$ in the generalized Proca theory for an Abelian vector field, which contains a coupling between two gauge field strength tensors and the double dual Riemann tensor, is not constrained as it does not alter the gravitational waves speed \cite{Baker:2017hug} \footnote{This is in contrast to the scalar Galileon case where a non-trivial configuration of the field and a coupling to the Weyl tensor are a sufficient condition to generate an anomalous speed for the gravitational waves \cite{Bettoni:2016mij}.}.  The harmful terms seem then to be the couplings with the Ricci scalar.  Thus, an adequate relation between the $\alpha$ and $\kappa$ parameters in $\alpha \mathcal{L}_4^1 + \kappa \mathcal{L}_4^2 \subset \mathcal{L}_4$ could deactivate the apparently harmful effect of the couplings with the Ricci scalar, at least for the cosmic triad configuration, making of this model a phenomenologically viable alternative.  Of course, if this is possible, the cosmological implications of $\alpha \mathcal{L}_4^1 + \kappa \mathcal{L}_4^2$ must be studied having in mind that the dark energy mechanism presented in this paper might not be counterbalanced by $\kappa \mathcal{L}_4^2$ \footnote{It is worth mentioning that self-accelerating cosmologies in the scalar Galileon case are incompatible with observations \cite{Lombriser:2015sxa,Brax:2015dma,Arai:2017hxj}.  It remains to be seen if generalized Proca theories have the same fate.}.  We expect to address this issue in a forthcoming publication.

{\it Acknowledgements} - This work was supported by the following grants: Colciencias - 123365843539 CT-FP44842-081-2014, DIEF de Ciencias - UIS - 2312, VCTI - UAN - 2017239, and Centro de Investigaciones - USTA - GIDPTOBASIC-UISUANUVALLEP12017.  We want to acknowledge Alejandro Guarnizo, Luis Gabriel G\'omez, Fabio Duv\'an Lora, C\'esar Alonso Valenzuela, Carlos Mauricio Nieto, and Miguel Zumalac\'arregui for useful discussions and help at different stages of this project.  Y. R. wants to give a special acknowledgement to Erwan Allys and Patrick Peter, who the generalized $SU(2)$ Proca theory was constructed with, for very instructive discussions, valuable help, and for sharing some research objectives.

\bibliography{bibli.bib} 

\end{document}